# Ultrafast Quasiparticle Dynamics and Electron-Phonon Coupling in $(Li_{0.84}Fe_{0.16})OHFe_{0.98}Se$

Q. Wu[1,2], H. X. Zhou[1], Y. L. Wu[1], L. L. Hu[1], S. L. Ni[1,2], Y. C. Tian[1], F. Sun[1,2], F. Zhou[1], X. L. Dong[1], Z. X. Zhao[1], and Jimin Zhao[1,2,3,*]

[1] *Beijing National Laboratory for Condensed Matter Physics, Institute of Physics, Chinese Academy of Sciences, Beijing 100190, China*

[2] *School of Physical Sciences, University of Chinese Academy of Sciences, Beijing 100049, China*

[3] *Songshan Lake Materials Laboratory, Dongguan, Guangdong 523808, China*

\* Corresponding author, Email: <u>jmzhao@iphy.ac.cn</u>

**Distinctive superconducting behaviors between bulk and monolayer FeSe make it challenging to obtain a unified picture of all FeSe-based superconductors. We investigate the ultrafast quasiparticle (QP) dynamics of an intercalated superconductor $(Li_{1-x}Fe_x)OHFe_{1-y}Se$, which is a bulk crystal but shares a similar electronic structure with single-layer FeSe on $SrTiO_3$. We obtain the electron-phonon coupling (EPC) constant $\lambda_{A_{1g}}$ (0.22 ±0.04), which well bridges that of bulk FeSe crystal and single-layer FeSe on $SrTiO_3$. Moreover, we find that such a positive correlation between $\lambda_{A_{1g}}$ and superconducting $T_c$ holds among all known FeSe-based superconductors, even in line with reported FeAs-based superconductors. Our observation indicates possible universal role of EPC in the superconductivity of all known categories of iron-based superconductors, which is a critical step towards achieving a unified superconducting mechanism for all iron-based superconductors.**

Most high-temperature superconductors exhibit quasi-two-dimensional superconducting (SC) layers. However, a unified physical picture of the FeSe-based [1] and FeAs-based superconductors [2] remains elusive thus far, regardless of the common layered feature they share. The recent discoveries of single-layer FeSe (on $SrTiO_3$ or similar substrates) interface superconductors [3] and $(Li_{1-x}Fe_x)OHFe_{1-y}Se$ intercalated superconductors [4] seemingly reinforced this uncertainty. However, by examining these newly discovered crucial types of superconductors, possible connections among all categories of iron-based superconductors can be uncovered, leading towards a universal picture of them all.

In this Letter, we investigate the electron-phonon coupling (EPC) strength in these superconductors by using ultrafast spectroscopy [5-9] to gain insights into a unified understanding of iron-based superconductors. Ultrafast spectroscopy of iron-based superconductors has been previously reported extensively [5,6,10-12], mainly identifying the SC transition and obtaining the SC parameters. Here we focus on the EPC and make a systematic investigation of its relationship to superconductivity. Specifically, we probe the excited state quasiparticle (QP) dynamics of a bulk crystal $(Li_{0.84}Fe_{0.16})OHFe_{0.98}Se$ [13]. This technique allows direct observation of the QP lifetime, uncovering the phenomena above the SC gap and across the SC transition. Light pulses break the Cooper pairs and generate excited-state QPs, which recombine with a lifetime determined mainly by the EPC [5,6]. The stronger the EPC, the faster the energy transfer through EPC, and hence, the shorter the lifetime. We determine the EPC strength to be $\lambda_{A_{1g}} = 0.22 \pm 0.04$, which is larger than that of bulk FeSe ($\lambda_{A_{1g}} = 0.16$)

[5] and smaller than that of single-layer FeSe/SrTiO$_3$ ($\lambda_{A_{1g}} = 0.48$) [6].

(Li$_{1-x}$Fe$_x$)OHFe$_{1-y}$Se materials [4,13,14] consist of staggered stack of Fe$_{1-y}$Se and (Li$_{1-x}$Fe$_x$)OH layers; this structure is analogous to that of K$_x$Fe$_2$Se$_2$ [15], but with the *P*4*/nmm* space group instead of the *I*4*/m* space group. The electronic band structure of (Li$_{1-x}$Fe$_x$)OHFe$_{1-y}$Se is similar to that of single layer FeSe/SrTiO$_3$ [16] and bulk Rb$_x$Fe$_{2-y}$Se$_2$ [17]. As a result, (Li$_{1-x}$Fe$_x$)OHFe$_{1-y}$Se forms a key bridge between bulk FeSe, $A_x$Fe$_{2-y}$Se$_2$ ($A$ = K, Rb, Cs, Tl/K), and single-layer FeSe/SrTiO$_3$ superconductors. To some extent, (Li$_{1-x}$Fe$_x$)OHFe$_{1-y}$Se provides an excellent platform for gaining insights into the shared SC origin of these materials [13,16-18].

In this Letter, we demonstrate that the $\lambda_{A_{1g}}$ value of (Li$_{1-x}$Fe$_x$)OHFe$_{1-y}$Se bridges those of bulk FeSe, Fe$_{1.02}$Se$_{0.2}$Te$_{0.8}$, Fe$_{1.05}$Se$_{0.2}$Te$_{0.8}$, K$_x$Fe$_2$Se$_2$ and single-layer FeSe/SrTiO$_3$. Significantly, by examining this and the reported $\lambda_{A_{1g}}$ values measured using ultrafast spectroscopy for many FeSe-based and FeAs-based superconductors, as well as those calculated in theory, we discover a positive correlation between the EPC strength $\lambda_{A_{1g}}$ and the SC transition temperature $T_c$. Our results indicate the possible important role of EPC in the superconductivity for all categories of iron-based superconductors, thus contributing to a unified understanding of the SC mechanism.

Ultrafast laser pulses with a central wavelength of 800 nm, 70 fs pulse width and 250 kHz repetition rate are used to excite and probe the dynamics of QPs, as well as to generate and detect coherent phonons, in a (Li$_{0.84}$Fe$_{0.16}$)OHFe$_{0.98}$Se single crystal. The time-resolved pump-probe weak detection experimental setup is similar to that of Ref. [6], with the polarizations of the pump and probe beams being perpendicular to each other to enhance the signal-to-noise ratio. Moreover, a balanced detector is used to further improve the signal. When the pump pulses arrive at the sample, a very small portion of the paired (if in the SC state) electrons are promoted to the excited states far above the SC gap, forming QPs that relax gradually thereafter to the ground state.

Our sample is synthesized via a hydrothermal ion-exchange technique, the details of which are described in Ref. [13]. The measured sample is $1.2 \times 2$ mm$^2$-sized, composing of micron-sized single-crystal flakes, and is kept in a refrigerator to prevent degradation before and after the experiments. Its surface is not entirely optical flat, resulting in diffusive reflections, which makes collecting the reflected beam and further obtaining useful signals challenging.

Figure 1(a) shows the differential reflectivity $\Delta R/R$ of our (Li$_{0.84}$Fe$_{0.16}$)OHFe$_{0.98}$Se sample (data are offset for clarity), with the schematic crystal structure illustrated in Fig. 1(b). The value of $\Delta R/R$ reflects the transient change of the refractive index, for which the electronic and lattice contributions can be clearly separated [19]. In Fig. 1(a), $\Delta R/R$ is proportional to the density of the excited state QPs. We color-map the complete set of data as a function of both temperature and time delay in Fig. 1(c) (on the logarithmic scale for clarity). The color change reflects the evolution of the QP density, which exhibits a phase transition at 40 K. This color change is similar to that for the SC transition in the monolayer FeSe superconductor [6] and many others [10]. The phase transition temperature 40 K is also similar to the reported SC $T_c$ for (Li$_{1-x}$Fe$_x$)OHFe$_{1-y}$Se by other experimental methods [4,13,14,18].

To further verify that the color change represents a SC transition, we quantitatively analyze the excited-state QP dynamics. The data in Fig. 1(c) can be fitted very well by using two dynamical components, as $\Delta R/R = A_{fast}\exp(-t/\tau_{fast}) + A_{slow}\exp(-t/\tau_{slow}) + y_0$, where $A_{fast}$ ($A_{slow}$) and $\tau_{fast}$ ($\tau_{slow}$) are the amplitude and lifetime of the fast (slow) component, respectively. For details of the data fitting, see Supplemental Material [20]. The fast relaxation component $A_{fast}\exp(-t/\tau_{fast})$ mainly reflects the EPC, while the slow component $A_{slow}\exp(-t/\tau_{slow})$ mainly reflects the phonon-phonon scattering. These two components together depict the temporal density evolution of the excited-state QPs. These QPs form only a very small portion of the electrons condensed into ground-state Cooper pairs [20]. The third term $y_0$ is a constant within the scanning range and slightly varies with temperature. Indeed, it might be a component with a lifetime much longer than the scanning range [5,46], which could be due to scatterings among acoustic phonons. We mainly focus on the fast and slow components.

We investigate the temperature dependence of the QP dynamics. Figure 2(a) shows that $A_{fast}$ behaves similar to that of single-layer FeSe/SrTiO$_3$ [6]. Figure 2(b) demonstrates that $\tau_{fast}$ is nearly a constant for the whole temperature range. Figure 2(c) exhibits that $A_{slow}$ experiences a dramatic decrease near 39.5 K, similar to those in monolayer FeSe/SrTiO$_3$ [6] and other superconductors [47]. Figure 2(d) illustrates that $\tau_{slow}$ forms a peak (*i.e.*, enhanced QP lifetime) at 39.5 K, directly manifesting the well-known phonon bottleneck effect. Microscopic detailed balancing between the excited-state QPs and the high-frequency phonons (abbreviated as HFPs, defined by $E > 2\Delta(T)$) dictates the cross-gap recombination of the QPs [6,11,47]. The photo-excited QPs relax to the ground state to form Cooper pairs, accompanied by releasing of HFPs. Conversely, the HFPs can break the Cooper pairs, generating excited-state QPs. Only when HFPs decay or propagate away, will the detailed balance break [20], thus leading to longer QP lifetimes owing to more phonons fulfilling the HFP criteria near $T_c$ (due to the SC gap closing). The simultaneous observation of the abrupt decrease in the amplitude [Fig. 2(c)] and the lifetime maximum [Fig. 2(d)] at the same temperature indicates a SC phase transition [6,47]. To further verify this, we quantitatively analyze the temperature dependence of $A_{slow}$ and $\tau_{slow}$ [20] by using the Rothwarf-Taylor (RT) model [48] and its extended derivation [49]. The RT model was initially developed to understand the superconducting behavior from the dynamical aspect. It describes the detailed balance between the excited state QPs and the HFPs, involving the Cooper pairs. The RT model is usually applicable when the weak detection condition [20] is fulfilled. The curves based on this model for a superconductor are plotted in Figs. 2(c) and 2(d) (red solid curves), respectively. Both model curves reveal that the $T_c$ is 39.7 ± 0.5 K and the SC gap $\Delta(0)$ is 14.3 ± 1.2 meV. The density of thermally excited QPs, $n_T$, is also illustrated in the inset of Fig. 2(c). We thus identify $\Delta(0)/k_B T_c$ to have a value of 4.1, which is much larger than the BCS prediction $\Delta(0)/k_B T_c = 1.76$. Here the evidence and parameters of the SC transition are obtained by probing the excited-state ultrafast dynamics, rather than by measuring the static near-Fermi-surface properties. The results in Figs. 1 and 2 are obtained within the weak detection regime. By weak detection, we propose two different levels of criteria: (A) no thermal effect and (B) no prominent destruction of the SC component, which are both fulfilled in our experiment [20, see

Sections 3 and 4].

In the following paragraphs, we investigate the EPC. By measuring the QP ultrafast dynamics under relatively high excitation laser fluence, the EPC constant $\lambda$ can be directly obtained. This is by using a phenomenological model by Allen [50], which elucidates the relation between the QP relaxation rate and the EPC strength. It can be applied to both superconductors and many other solids. In the model, the thermal relaxation rate $\gamma_T$ of the QPs is determined jointly by the EPC constant $\lambda$ and the electron temperature $T_e$ in a relation , where   is the second moment of the Eliashberg function, $T_L$ is the lattice temperature, and $\Omega$ is the phonon frequency. We measure the electron temperature $T_e$ and the QP relaxation rate $\gamma_T$ in our ultrafast dynamics experiment.

To obtain $T_e$, note that the high order terms in the equation of $\gamma_T$ are negligible only when $T_e T_L$ is relatively large, *i.e.*, when $(E_{phonon})^2 \leq E_{\text{excited-state electron}} \cdot E_{\text{ground-state electron}}$, which is fulfilled when the excited-state electron energy is much higher than the optical phonon energy. This condition is fulfilled in high-fluence excitation, which is implemented in our experiment, with a pump fluence of $F = 2.2$ mJ/cm$^2$. We calculate $T_e$ using the relation , where $l_s$ is the optical penetration depth, and $\kappa_v = \kappa \rho_n$ is the electronic heat capacity coefficient for unit volume, with $\kappa$ being the electronic heat capacity coefficient (Sommerfeld parameter) and $\rho_n$ being the mole density [20]. By interpreting $\kappa = 100$ mJ/(mol·K$^2$) from Ref. [4] and taking $T_L = 7$ K, we obtain $T_e = 548$ K. Because $T_L$ only affects $T_e$ less than a few Kelvin (when $T_L$ is relatively low), $T_e$ is close to a constant [20].

By definition, $\gamma_T = \left| \dfrac{1}{T_e - T_L} \dfrac{d(T_e - T_L)}{dt} \right|_{t \to 0}$ [50]. It is known that the difference in electron and lattice temperatures is proportional to the differential reflectivity, as ($T_e$ - $T_L$)∝$\Delta R/R$ [51]. Furthermore, at $t\to 0$, the fast component dominates. Thus we have $\gamma_T = \left|\frac{1}{\Delta R}\frac{d(\Delta R)}{dt}\right|_{t\to 0} \approx 1/\tau_{fast}$. Hence, we can directly measure $\gamma_T$ through $\tau_{fast}$ in our time-resolved ultrafast spectroscopy experiment. A typical ultrafast dynamical response under such high fluence is shown in Fig. 3(a). Fitting the electronic dynamics (after subtracting the overall dynamics of phonon oscillations) at a low temperature ($T_L$ = 7 K), we obtain after deconvolution that $\tau_{fast}$ = 0.34 ± 0.05 ps [20], where $\tau_{fast}$ is nearly identical for all the low temperature values. This lifetime is longer than the 0.23 ps lifetime of single-layer FeSe/SrTiO$_3$ system [6] and shorter than the 1.75 ps lifetime of bulk FeSe [5]. Note that in the high fluence regime $\tau_{fast}$ increases with fluence, whereas the value of $\lambda$ remains constant [20]. With the above observed values of $T_e$ and $\gamma_T$ we derive that $\lambda\langle\Omega^2\rangle$ = 2.34×10$^{26}$ Hz$^2$ (*i.e.*, 101 meV$^2$).

The phonon information is needed before obtaining $\lambda$. We generate and detect coherent phonons [52-55] in (Li$_{0.84}$Fe$_{0.16}$)OHFe$_{0.98}$Se with the same high pump (2.2 mJ/cm$^2$) and probe (0.28 mJ/cm$^2$) fluences, as shown in Fig. 3(a). We extract the coherent oscillation and perform Fourier transformation to obtain its frequency [Fig. 3(a) inset], which is 5.11 THz (*i.e.*, 21.2 meV or 171 cm$^{-1}$). We assign it to be the A$_{1g}$ mode of the Se atoms, which corresponds to layer breathing along the *c*-axis; and ascribe its generation mechanism to be displacive excitation of coherent phonons [20]. The temperature dependence of the coherent phonon is investigated [Figs. 3(b-e)], which exhibits a regular anharmonic phonon decay [20,52] [Fig. 3(e)]. Figures 3(c-e) show that no structural phase transition is observed in the temperature range of 100-

180 K. The smoothness of our coherent phonon data also confirm the reliability of all our varying-temperature experiments, including those for the QPs (for example, there is no prominent sample motion, and the sample flake size turns out does form an actual obstacle by our detection scheme).

Note that demonstrating the $A_{1g}$ phonon does not indicate that the $A_{1g}$ mode phonon is exactly the pairing glue. So far, there is no consensus on which phonon mode contributes most to the EPC strength $\lambda$. If we use the above measured $A_{1g}$ phonon frequency, 21.2 meV, we only obtain a nominal EPC constant $\lambda_{A_{1g}} = 0.22 \pm 0.04$ [20], instead of the EPC $\lambda$. In fact, nominal values of $\lambda_{A_{1g}}$ in superconductors are frequently reported, however, in the name of $\lambda$ [5,6,9,56,57]. Here, we clearly distinguish $\lambda_{A_{1g}}$ and $\lambda$, which clearly illustrates important basic facts that are otherwise mixed, concealing important conclusions. We found that (i) by distinguishing $\lambda_{A_{1g}}$ and $\lambda$, we are now able to compare better the (many reported) experimental results with theoretical calculations; (ii) the values of $\lambda_{A_{1g}}$ are all based on experimental facts, thus deserve consideration and comparison; and (iii) significantly, both $\lambda_{A_{1g}}$ and $\lambda$ exhibit positive correlations with $T_c$, but not when $\lambda_{A_{1g}}$ and $\lambda$ are mixed. The values of $\lambda_{A1g}$ and $\lambda$ can be related (converted into each other) through the relation $\lambda_{A1g} = \lambda()$, which we derive based on Refs. [50,58,59]: Taking the first term of the equation by Allen [50], which contains the relation $\lambda \propto 1/<\Omega^2>$ by McMillan [59], we define $\lambda<\Omega^2> \equiv \lambda_{A_{1g}}<\Omega_{A_{1g}}^2> = \lambda_{A_{1g}}\Omega_{A_{1g}}^2$, where $\lambda_{A_{1g}}$ is the nominal EPC strength corresponding to the $A_{1g}$ mode phonon. Further, in theoretical calculations, it is convention to take the Allen-Dynes treatment, where logarithmic phonon frequency $\Omega_{\log}$ is taken to be the overall average [58], thus we write

$\lambda<\Omega^2> = \lambda<\Omega_{\log}^2> = \lambda\Omega_{\log}^2$. Combining the two, we retrieve the values of $\lambda_{A_{1g}}$ by using $\lambda_{A_{1g}} = \lambda(\Omega_{\log}^2/\Omega_{A_{1g}}^2)$.

The EPC has been speculated to play an active role in certain individual types of iron-based high-$T_c$ superconductors, particularly such as the single-layer system [3,6,60,61]. Here, instead, we contemplate on *all* categories of iron-based superconductors. We plot the available $T_c$ versus $\lambda_{A_{1g}}$ values for different optimally doped iron-based superconductors in Fig. 4. To avoid systematic discrepancies due to different experimental methods, we mainly summarize $\lambda_{A_{1g}}$ values obtained by using time-resolved ultrafast optical spectroscopy (including time-resolved angle-resolved photoemission spectroscopy, red spheres and squares) [5,6,8,9,57,62], as well as theoretical results (blue spheres and squares) [63-69]. A result from angle-resolved photoemission spectroscopy (golden sphere) is also included [60]. Note that some of the data are retrieved by the authors [20]. Furthermore, we perform ultrafast optical spectroscopy experiments on $Fe_{1.05}Te_{0.8}Se_{0.2}$ ($T_c = 10$ K) and $Fe_{1.01}Te_{0.8}Se_{0.2}$ ($T_c = 13.5$ K) single crystals, respectively. The experimental and analysis details are similar, and we obtain the $\lambda_{A_{1g}}$ values to be 0.11 and 0.13, respectively [20]. All the experimental results in Fig. 4 are (were) obtained by 800 nm laser beam excitations (except for bulk FeSe [5] with 400 nm excitation). Because the QP lifetime is nearly constant at relatively high energy [70], it is fully reasonable to make a comparison among these results.

Significantly, all the data points are located within a purple stripe region centered on a pink curve in Fig. 4, indicating a positive correlation between $T_c$ and $\lambda_{A_{1g}}$. Here the pink curve is a revised version of Allen-Dynes formula [58], based on $\lambda_{A1g}$ and $\Omega_{A1g}$ [20]. The Allen-Dynes formula reveals a positive correlation between $\lambda$ and the SC $T_c$, and it compares well with the experimental data mostly for conventional superconductors. Similarly, we also plot available reported theoretical results of $\lambda$ in Fig. 4 (marked by light gray squares or spheres), which are located within a brown stripe region centered on an orange curve (another revised version of Allen-Dynes formula, based on $\lambda$ and $\Omega_{\log}$ [20]. Because $\lambda_{A1g}$ and $\lambda$ are explicitly distinguished here, and EPC is considered systematically for the whole family of iron-based superconductors, we are able to establish a hitherto unrecognized correlation between $T_c$ and $\lambda_{A1g}$ ($\lambda$) for iron-based superconductors, using all the available theoretical and experimental results. We emphasize that the positive correlation itself, rather than the details of the fitting equations, is the most significant discovery.

Remarkably, bulk, single-layer, and intercalated iron-based superconductors all obey these two correlations; and both FeSe- and FeAs-based superconductors obey the same correlations as well [Fig. 4]. This observation strongly suggests that a shared SC origin is ubiquitous in all categories of iron-based superconductors, whereby the EPC plays a universal role. We speculate that such EPC may not be a conventional type of EPC. Moreover, we do not rule out other origins in making important contributions to the SC mechanism (for instance, spin interactions or charge modulations may affect the EPC through modifying the electronic states or the phonons [20]).

In summary, we investigate the QP dynamics of $(Li_{0.84}Fe_{0.16})OHFe_{0.98}Se$ using time-resolved ultrafast spectroscopy. By the temperature dependence measurements, we identify the SC properties, such as the SC $T_c$ and $\Delta$ values. More importantly, by the QP lifetime, we obtained the EPC strength $\lambda_{A1g}$ of $(Li_{0.84}Fe_{0.16})OHFe_{0.98}Se$, along with the $\lambda_{A1g}$ for $Fe_{1.05}Te_{0.8}Se_{0.2}$ and $Fe_{1.01}Te_{0.8}Se_{0.2}$ by our additional experiments. Significantly, we discover a positive correlation between $T_c$ and the EPC strength, either in the form of $\lambda_{A1g}$ or $\lambda$. Our finding demonstrates the likely crucial role of phonon in the high temperature SC mechanism for iron-based superconductors. Our experimental result suggests the possible existence of a unified framework for understanding iron-based superconductors, including the monolayer system. Our investigation is a critical step towards a unified superconducting mechanism for all iron-based superconductors.

**Acknowledgments**

This work was supported by the National Key Research and Development Program of China (Grants No. 2017YFA0303603, 2016YFA0300300), the National Natural Science Foundation of China (Grants No. 11574383, 11774408, and 11574370), the Frontier Program of the Chinese Academy of Sciences (Grant No. QYZDY-SSW-SLH001), the Strategic Priority Research Program of CAS (Grant No. XDB30000000), Beijing Natural Science Foundation (Grant No. 4191003), the International Partnership Program of Chinese Academy of Sciences (Grant No. GJHZ1826), and the CAS Interdisciplinary Innovation Team.

**Figures and captions**

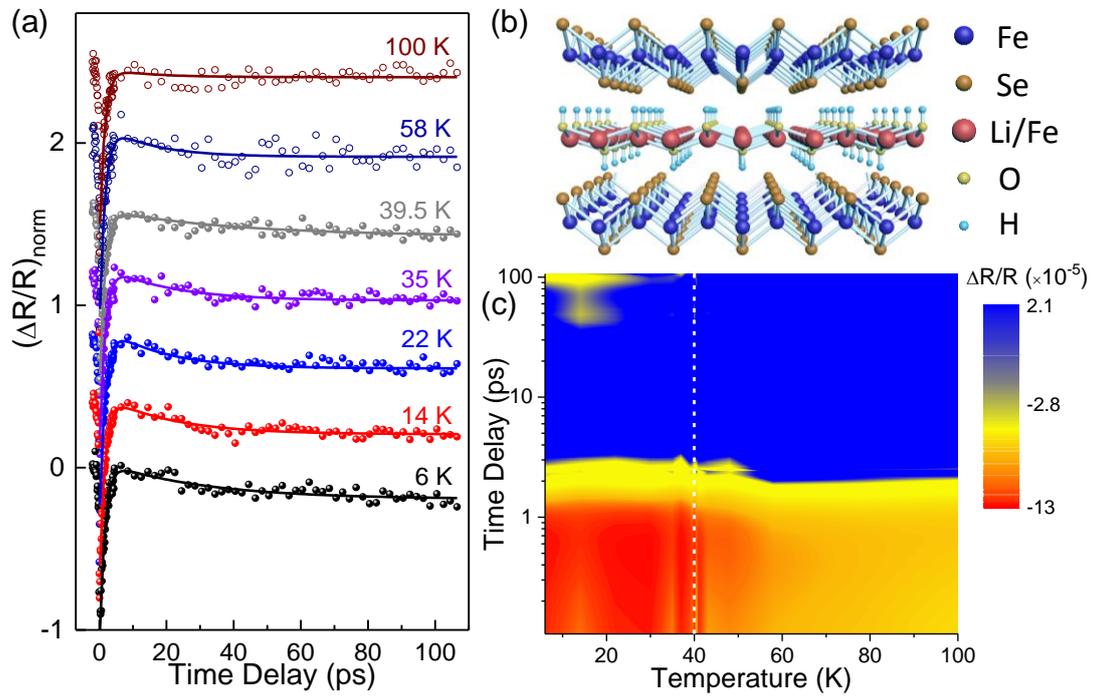

FIG. 1. Ultrafast QP dynamics in $(Li_{0.84}Fe_{0.16})OHFe_{0.98}Se$. (a) Normalized time-resolved differential reflectivity $\Delta R/R$ at different temperatures (offset for clarity). Spheres: scanning data below SC $T_c$; open circles: scanning data above SC $T_c$; solid curves: 2-exponential fittings. (b) Schematic lattice structure with $P4/nmm$ symmetry. (c) Colormap of $\Delta R/R$ as a function of both the probe beam delay time and temperature. The color change at 40 K reveals a likely SC phase transition (indicated by a white dashed line).

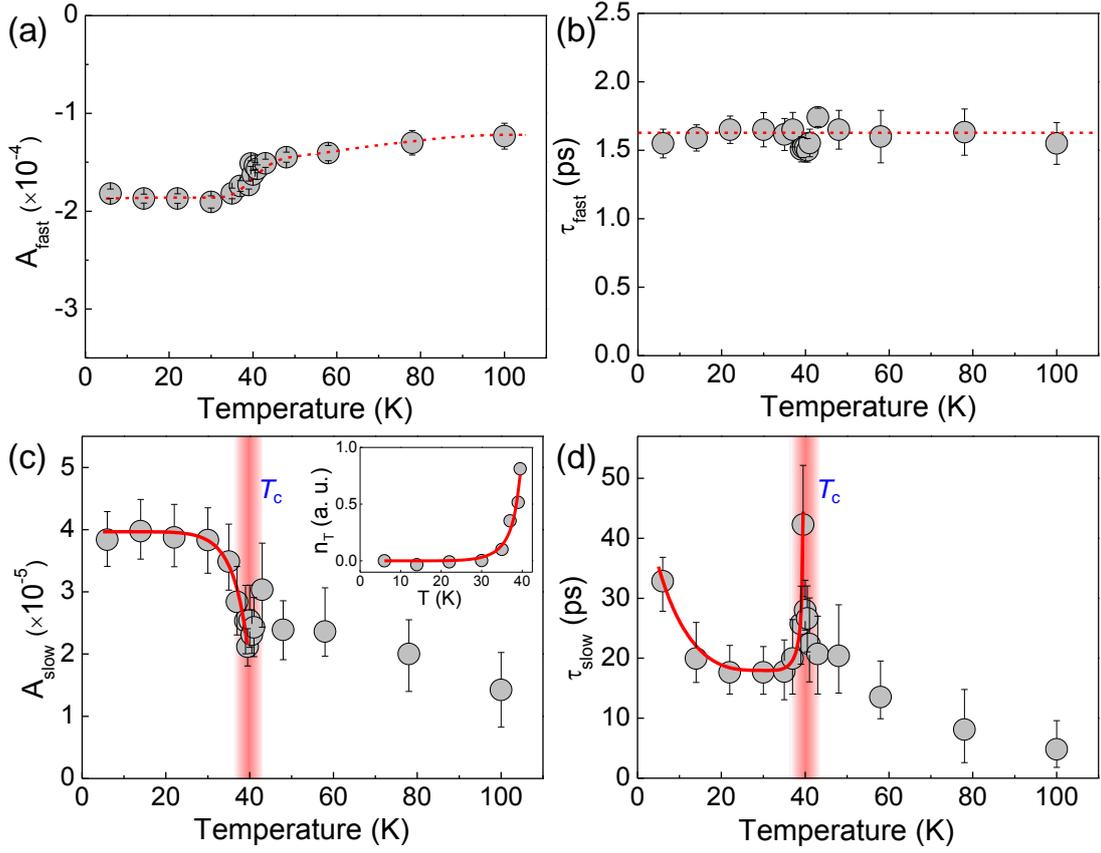

FIG. 2. Quantitative QP dynamic evidence of the SC phase transition. (a,b) Amplitude $A_{fast}$ and lifetime $\tau_{fast}$ of the fast relaxation component and their temperature dependence. (c,d) Amplitude $A_{slow}$ and lifetime $\tau_{slow}$ of the slow relaxation component and their temperature dependence. Inset: thermal carrier density. Red solid curves in (c) and (d): theoretical data fitting based on the phenomenological Rothwarf-Taylor model. Dashed curves in (a) and (b): visual guides. Red vertical bars in (c) and (d): temperature region where SC phase transition occurs.

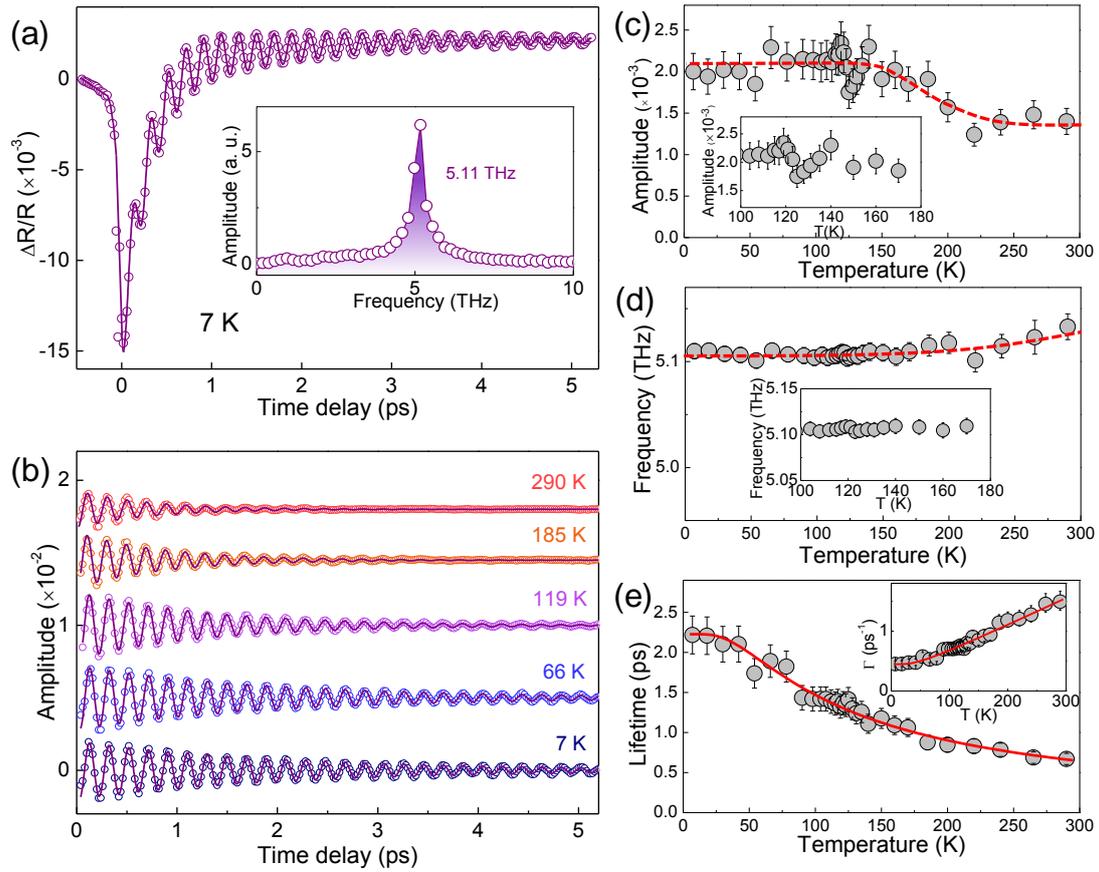

FIG. 3. Coherent optical phonon and its temperature dependence. (a) A typical time-resolved coherent optical phonon vibration at 7 K, which is superimposed on the electronic relaxation. Inset: the fast Fourier transform result showing a clear peak at 5.11 THz. The pump fluence is 2.2 mJ/cm$^2$. (b) Extracted phonon oscillations at several temperatures. (c-e) Temperature dependence of the phonon amplitude, frequency, and lifetime. Dashed curves: visual guides. Solid curves: anharmonic decay fitting results. Insets in (c-e): magnified views (c,d) and the corresponding phonon decay rate (e).

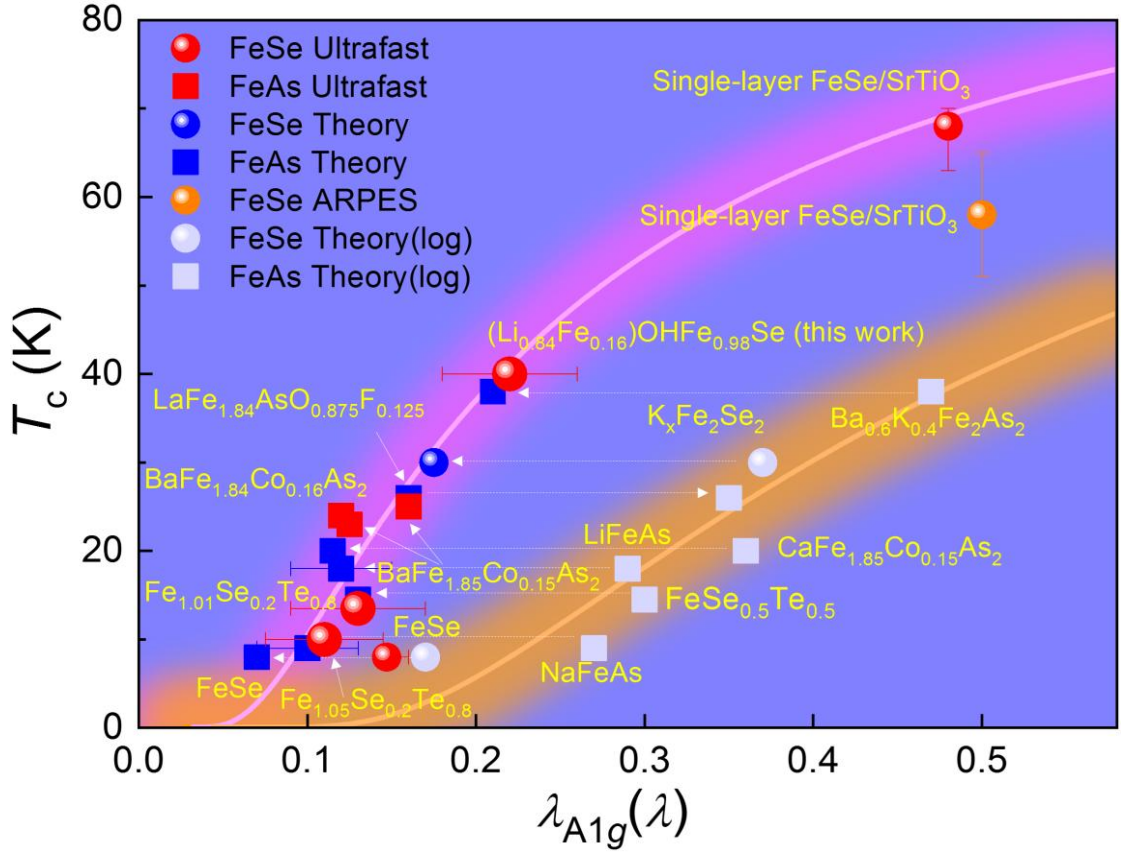

FIG. 4. Positive correlation between the EPC strength $\lambda_{A_{1g}}(\lambda)$ and SC $T_c$ in iron-based superconductors. The SC $T_c$ positively correlates to the EPC strength $\lambda_{A_{1g}}(\lambda)$. Red spheres (squares): ultrafast spectroscopy results of FeSe (FeAs)-based superconductors [three data from this work,5,6,8,9,57,62]. Golden sphere: angle-resolved photoemission spectroscopy result [60]. Blue sphere (squares): retrieved theoretical calculation results of FeSe (FeAs)-based superconductors [63-69]. Light gray spheres (squares): theoretical results of $\lambda$ for FeSe (FeAs)-based superconductors. Pink (orange) curve: ansatz modified version of Allen-Dynes formula using $\lambda_{A_{1g}}(\lambda)$ and $\Omega_{A_{1g}}(\Omega_{\log})$. White dashed arrows: indications of retrieving $\lambda_{A_{1g}}$ from $\lambda$, vise versa.

Supplemental Material

# Ultrafast Quasiparticle Dynamics and Electron-Phonon Coupling in (Li$_{0.84}$Fe$_{0.16}$)OHFe$_{0.98}$Se

Q. Wu[1,2], H. X. Zhou[1], Y. L. Wu[1], L. L. Hu[1], S. L. Ni[1,2], Y. C. Tian[1], F. Sun[1,2], F. Zhou[1], X. L. Dong[1], Z. X. Zhao[1], and Jimin Zhao[1,2,3],*

[1] *Beijing National Laboratory for Condensed Matter Physics, Institute of Physics, Chinese Academy of Sciences, Beijing 100190, China.*
[2] *School of Physical Sciences, University of Chinese Academy of Sciences, Beijing 100049, China.*
[3] *Songshan Lake Materials Laboratory, Dongguan, Guangdong 523808, China*
\* *Corresponding author, Email: jmzhao@iphy.ac.cn*

## 1. THE 2 RELAXATION COMPONENTS OF THE QP DYNAMICS

We carefully fit the data to understand the dynamics components. Tentative fittings using both 2- and 3-exponential functions are carried out for various low temperatures, as shown in Fig. S1. To see the near time zero region clearly, we also include the semi-log presentation in Figs. S1(e-h). Furthermore, to see the quality of data fitting, the goodness-of-fit results are also illustrated in Figs. S1(i-n). It can be seen from Figs. S1(a-h) that the 2-exponential model better fits the data, which is confirmed by the analysis shown in Figs. S1(i-n) (the $R^2$ value of 2-exponential fit is larger than that of 3-exponential fit). Although the 3-exponential curve provides a result close to the best fitting with the 2-exponential function, it yields a slow component with a lifetime longer than 300 ps, which may be systematic artifacts, considering the 110 ps scanning range, or anything related to thermal fluctuations. Thus, the QP dynamics consists of two relaxation components, each being an exponential decay. This result shows that our results are in accord with a two-temperature model (*i.e.*, the so-called 2 TM model).

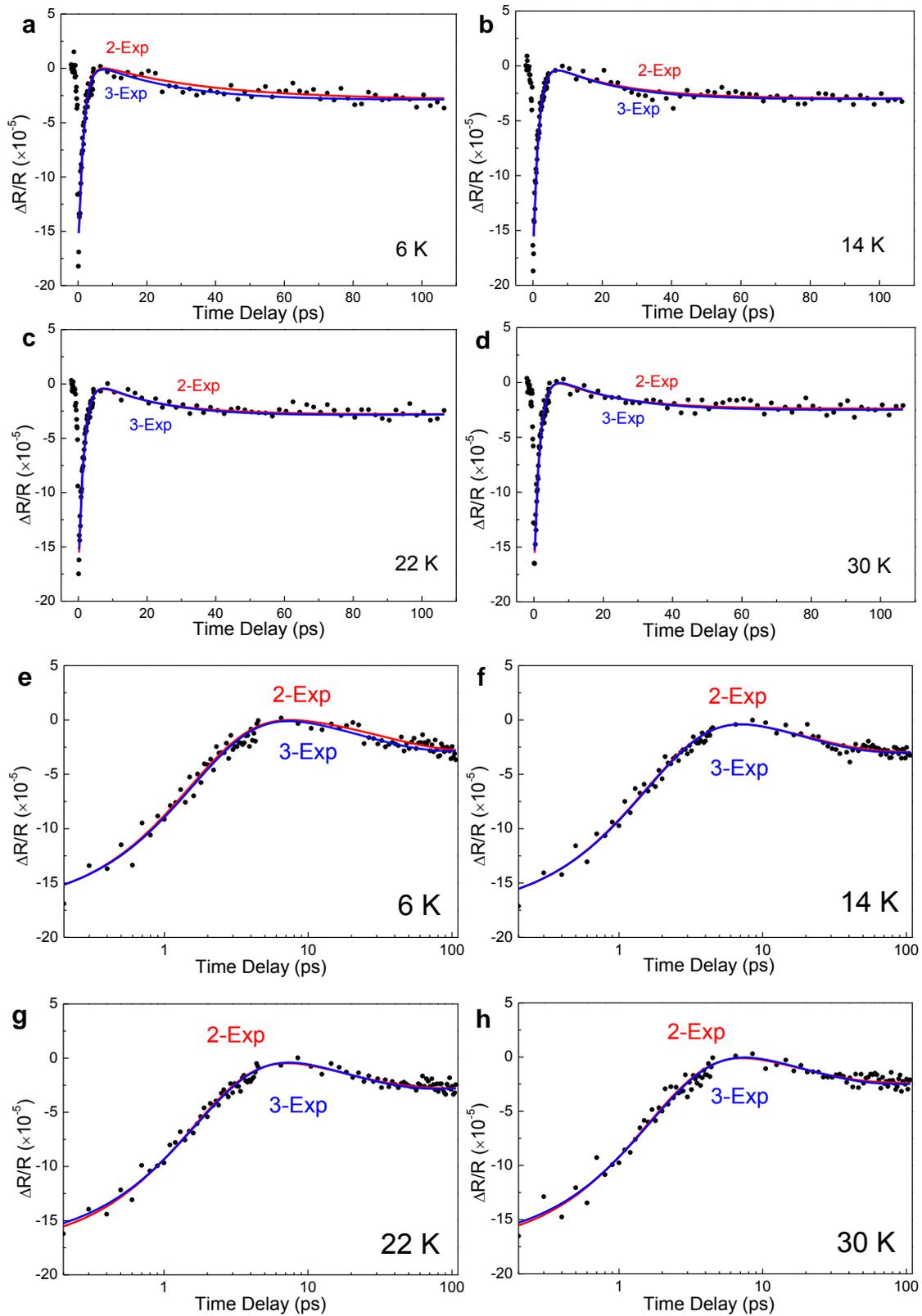

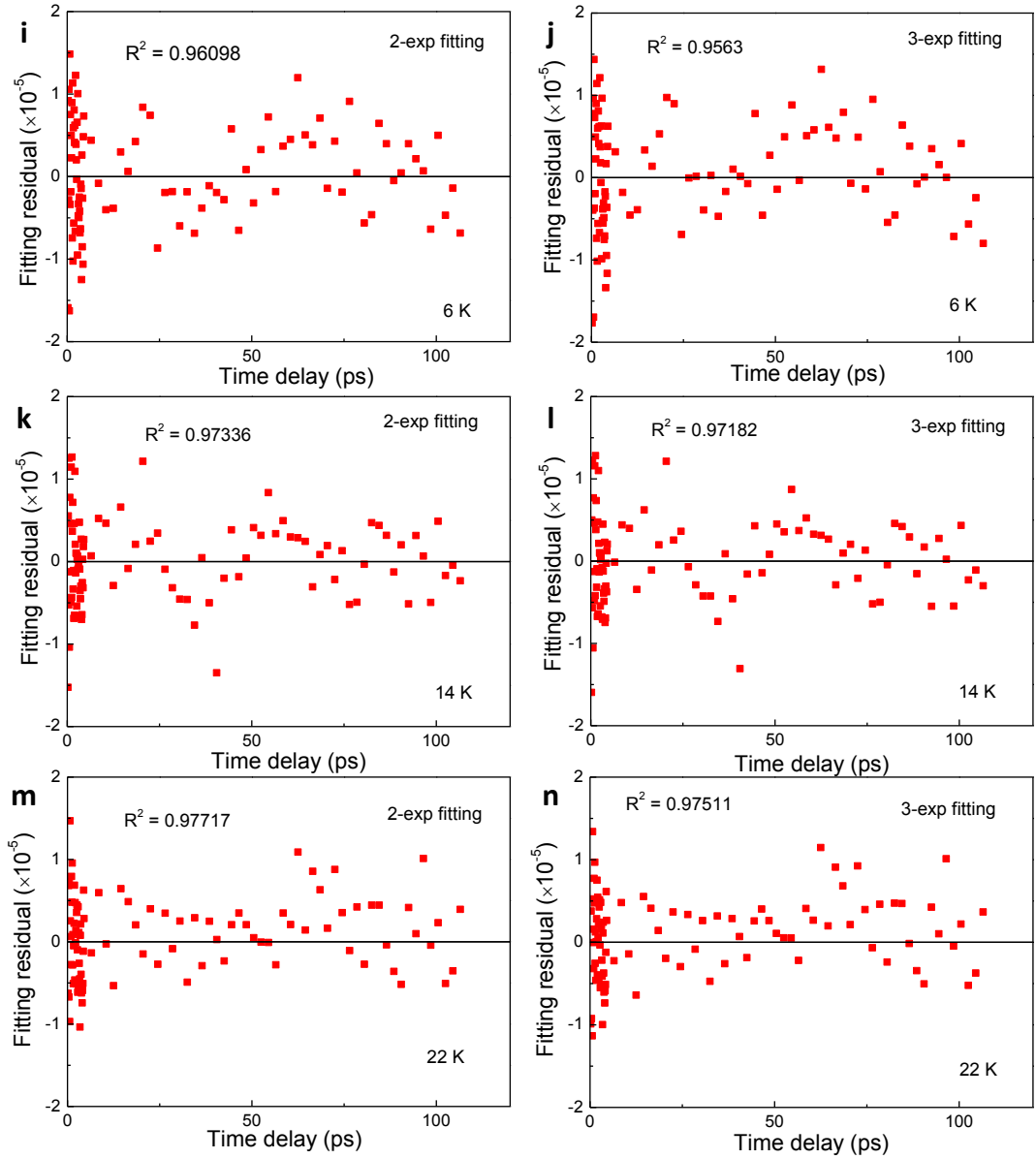

FIG. S1. Ultrafast dynamics results and the 2- and 3-exponential relaxation components. (a-d), Scanning traces at 6 K, 14 K, 22 K, and 30 K, respectively. Red and blue curves: fittings based on 2- and 3-exponential functions, respectively. (e-h), The semi-log plots of scanning traces at 6 K, 14 K, 22 K, and 30 K, respectively. (i-n), The residual and R square of 2- and 3-exponential fittings at 6 K, 14 K, and 22 K, respectively.

## 2. ROTHWARF-TAYLOR MODEL

The temperature dependence of the slow component can be fully explained by the

Rothwarf-Taylor model [1] and its significant extension derivations for ultrafast processes [2]. In the following, we provide details of the derivations.

The density of the thermally excited QPs $n_T$ has a temperature dependence of [3,4]

$$n_T \propto \sqrt{\Delta(T)T}\exp(-\Delta(T)/k_B T), \tag{1}$$

where $\Delta(T) = \Delta(0)\tanh\left(\Theta\sqrt{T_c/T - 1}\right)$ is the phenomenological SC gap [5,6], $k_B$ is the Boltzmann constant, $\Theta$ is the parameter which reflects the coupling strength (which is 1.77 for weak coupling case). Here, based on our data fitting, we take $\Theta$ to be 2.02, which reveals a strong coupling in the SC state.

The Rothwarf-Taylor equation reads [1]

$$\frac{dn}{dt} = I_0 + \eta N - \beta n^2, \tag{2}$$

$$\frac{dN}{dt} = J_0 - \frac{\eta N}{2} + \frac{\beta n^2}{2} - \Lambda(N - N_T), \tag{3}$$

where $n$ and $N$ are the density of QPs and high-frequency phonons (HFPs), respectively, $I_0$ and $J_0$ are the injection terms of QPs and HFPs, respectively, $\eta$ and $\beta$ are the Cooper pair breaking and formation coefficients, respectively, $\Lambda$ is the rate that phonons decay into low-energy phonons or propagate away from the active region, and $N_T$ is the density of thermally excited HFPs. Ignoring the injection terms ($I_0$ and $J_0$) and considering only the strong bottleneck case, we have at $t = t_s$

$$dn/dt = \eta N_s - \beta n_s^2 = 0, \tag{4}$$

where $t_s$ is the time when the QPs assume their maximum density. Taking $N(t = 0) = N_T + \Delta N$ and $n(t = 0) = n_T + \Delta n$ to be the number of HFPs and QPs at the excitation by pump pulse (neither for the ground state nor for the $t = t_s$ maximum QP density point), respectively, we obtain after simple derivation that, for $t = t_s$,

$$n_s = n_T + \frac{\eta}{4\beta} \frac{\frac{16\beta}{\eta}\Delta N + \frac{8\beta}{\eta}\Delta n}{\sqrt{\left(1+\frac{16\beta}{\eta}N_T+\frac{8\beta}{\eta}n_T\right)}}. \tag{5}$$

Our ultrafast dynamics experiment measures the differential reflectivity, $\Delta R/R$, which is proportional to the photo-carrier density. Explicitly, we have

$$A(T) \propto n_s - n_T = \frac{\eta}{4\beta} \frac{\frac{16\beta}{\eta}\Delta N + \frac{8\beta}{\eta}\Delta n}{\sqrt{\left(1+\frac{16\beta}{\eta}N_T+\frac{8\beta}{\eta}n_T\right)}}, \tag{6}$$

where $A(T)$ is the amplitude of differential reflectivity in ultrafast experiment. As $n_T|_{T=0\,K} = 0$ and $N_T \equiv (\beta/\eta) n_T^2$ for a steady state (see Eq. (2)), we have

$$A(0)/A(T) = 1 + 4\frac{\beta}{\eta}n_T. \tag{7}$$

Substituting Eq. (1) into Eq. (7), we can remove $\Delta N$ and $\Delta n$ in Eq. (6) to obtain

$$A_{slow}(T) \propto \left[\sqrt{\Delta(T)T}\exp\left(-\frac{\Delta(T)}{k_B T}\right) + C\right]^{-1}, \tag{8}$$

where $C$ is a constant containing $\eta$ and $\beta$, and we have replaced $A(T)$ by $A_{slow}(T)$. Equation (8) is the fitting equation used for Fig. 2(c) in the main text, where $\Delta(0)$ and $C$ are fitting parameters.

In parallel, regarding the lifetime, the ultrafast dynamics experiment directly measures the QP decay rate defined as $\tau^{-1} = \left|\frac{1}{n-n_T}\frac{dn}{dt}\right|_{t\to t_s}$, where $\tau$ is the lifetime of the slow component measured in our ultrafast time-resolved experiment, and $n$ is a function of temperature that approaches $n_s$ as $t$ approaches $t_s$. During the slow process, the photo-excited QPs relax to the ground state and form Cooper pairs, accompanied by releasing of HFPs. Conversely, the HFPs can break the Cooper pairs, generating excited-state QPs. Only when HFPs decay or propagate away, will the detailed balance break. The schematic plot is shown in Fig. S2 below.

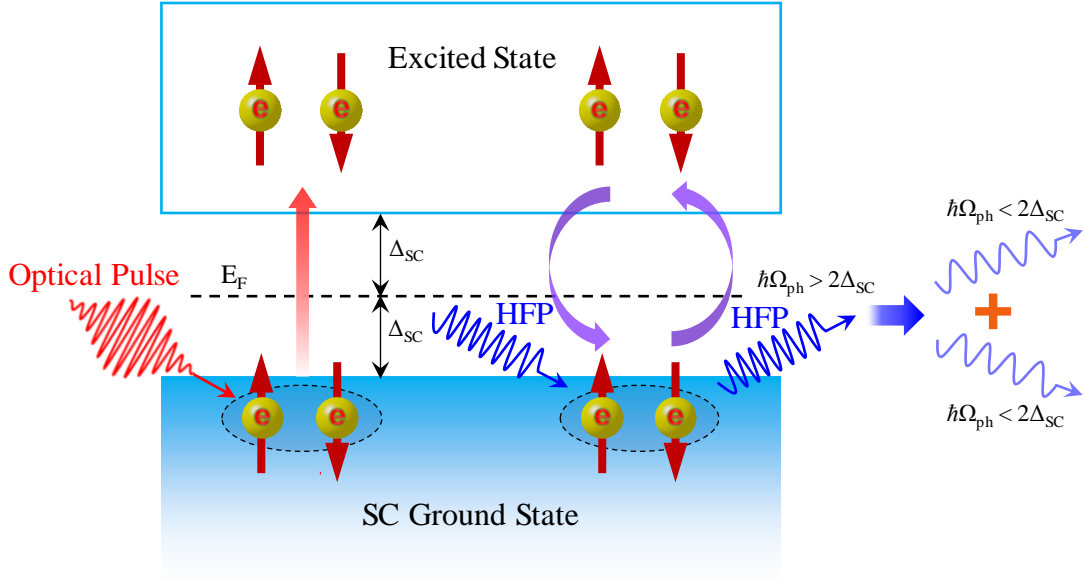

FIG. S2. Schematic of phonon-bottleneck effect in superconductors. Optical pulses break the Cooper pairs to form photo-excited QPs, which then relax and recombine to form Cooper pairs, releasing HFPs. Conversely, HFPs can break the Cooper pairs, generating excited-state QPs. This detailed balance between the QPs and HFPs only break when the HFPs decay into lower energy excitations or propagating away from the active region.

From Eq. (2,3), after performing simple mathematical operations, we obtain

$$\frac{dn}{dt} = -2\Lambda \frac{\beta}{\eta} \frac{(n^2 - n_T^2)}{\left(\frac{4\beta n_S}{\eta} + 1\right)}. \tag{9}$$

Assuming $4\beta n_S \ll \eta$, we derive the following:

$$\tau \approx \frac{\eta}{2\Lambda\beta(\delta + 2n_T)}. \tag{10}$$

The temperature dependence of $\Lambda$ depends on the magnitude of the SC gap as [4,7]

$$\Lambda \propto \Delta(T) + \alpha T \Delta(T)^4. \tag{11}$$

The first term corresponds to a phonon generating one lower-energy phonon (or propagating away from the active region), and the second term corresponds to a phonon generating two lower-energy phonons. Thus, the corresponding lifetime follows [7]:

$$\tau(T) \propto \{[\delta + 2n_T(T)][\Delta(T) + \alpha T\Delta(T)^4]\}^{-1}, \tag{12}$$

where $\Delta(0)$, $\delta$, and $\alpha$ are fitting parameters, and we have replaced $\tau(T)$ by $\tau_{\text{slow}}(T)$. Equation (12) is the fitting equation used for Fig. 2(d) in the main text. The abrupt reduction in $A_{\text{slow}}$ and prominent increase in $\tau_{\text{slow}}$ at approximately $T_c$ (Figs. 2c and d) are both attributed to the gradually vanishing SC gap with increasing temperature.

### 3. WEAK DETECTION CONDITION: (A) NO THERMAL EFFECT

By weak detection, we propose two different levels of criteria: (A) no thermal effect and (B) no prominent destruction of the SC component. For criterion (A), we have three aspects of verifications as follows.

#### 3.1 Linear fluence dependence range

We measure the fluence dependence of the QP dynamics, as shown in Fig. S3(a), with the interference artifact at the so-called time-zero removed for clarity. We also measure the fluence-dependent ultrafast dynamics at several typical temperatures with fluences ranging from 4.2 to 70 $\mu J/cm^2$ (Fig. S3). Note that in Fig. S3(a), we only show the 6 K result, whereas in Fig. S4, we show all the data for 6 K, 16 K, 28 K, 40.5 K, 43 K, and 60 K. We summarize the values of $|\Delta R/R|_{\text{max}}$ obtained in Fig. S3(b) (data are offset). At all temperatures, $|\Delta R/R|_{\text{max}}$ is proportional to the laser fluence, all with a slope of 1 (Fig. S3(b)).

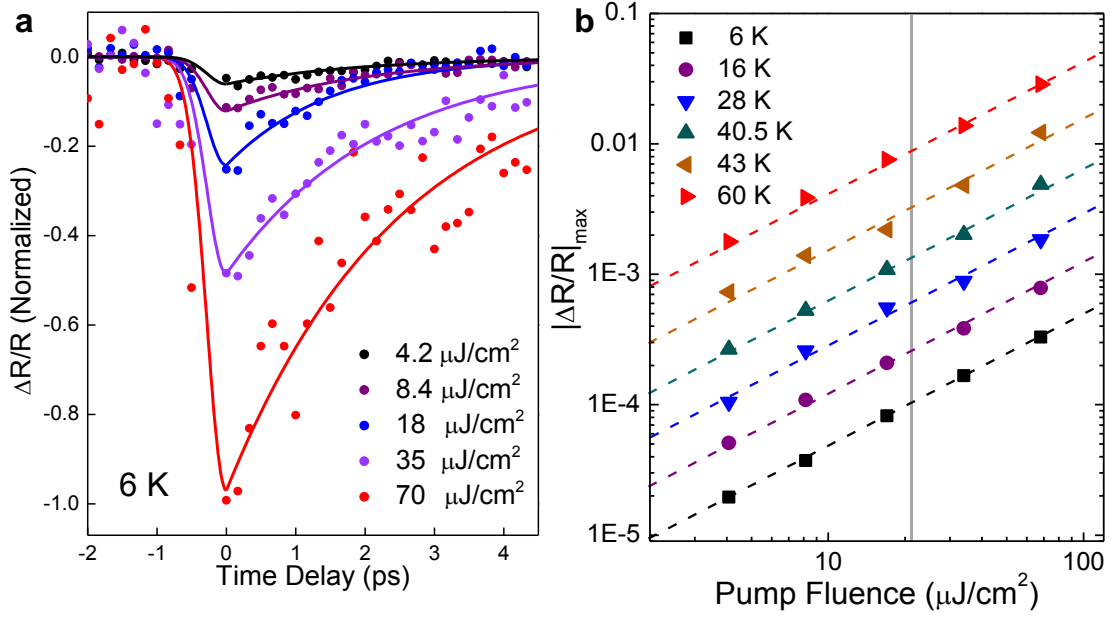

FIG. S3. Weak detection condition. (a) Fluence dependence of the differential reflectivity ($\Delta R/R$) at 6 K. Solid curves: visual guides. (b) Fluence dependence of $|\Delta R/R|_{max}$ obtained from (a) for various temperatures (offset for clarity). Dashed lines: fitting curves with a slope of 1. Gray vertical line: laser fluence used to obtain the QP dynamics data in Fig. 1 and 2 in the main text.

The fluence-dependent ultrafast dynamics that we measure (Fig. S3(a)) is also crucial in verifying that the laser pulse fluence used (21 $\mu J/cm^2$ for pump beam) for our temperature-dependent experiment fulfills the weak detection condition criterion (A). The results in Fig. S3(b) also demonstrate that for the dynamic range of fluence, from 4.2 to 70 $\mu J/cm^2$, the density of the photo-excited QPs is proportional to the pump fluence without saturation. Thus, the pump fluence we use (21 $\mu J/cm^2$) to obtain the data in Figs. 1 and 2 in the main text is appropriate, without breaking the SC condensed state, or introducing extra thermal effects [8].

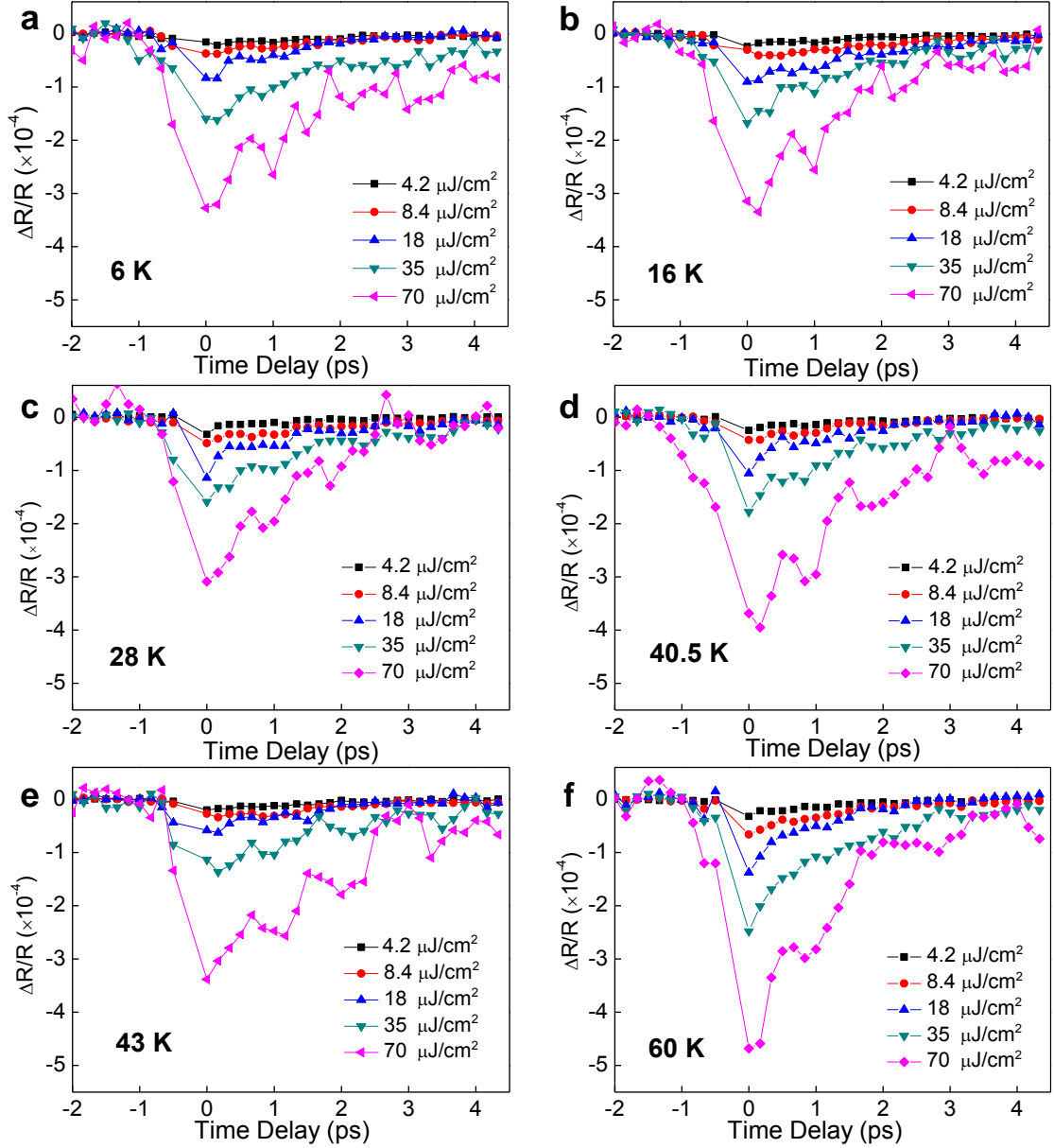

FIG. S4. Fluence dependence of the QP dynamics. (a-f) Fluence dependence of QP dynamics at 6, 16, 28, 40.5, 43 and 60 K, respectively. The fluences are 4.2, 8.4, 18, 35 and 70 $\mu J/cm^2$, respectively.

### 3.2 Less thermal effect with lower laser repetition rate

Different laser repetition rate affects the thermal effect dramatically. We compare our data with reported results (Fig. S5): it can be seen that, under the same single-pulse fluence, the thermal effect is much more prominent for higher repetition rate experiments (Fig. S6). With repetition rate increasing (from ① to ④), the ΔR/R saturation becomes more prominent, their corresponding saturation thresholds are >70 $\mu J/cm^2$, 10 $\mu J/cm^2$, 4.5 $\mu J/cm^2$, and 0.8 $\mu J/cm^2$, respectively, which is summarized in Fig. S6. It can be clearly seen that lower repetition rate allows for a much higher fluence threshold.

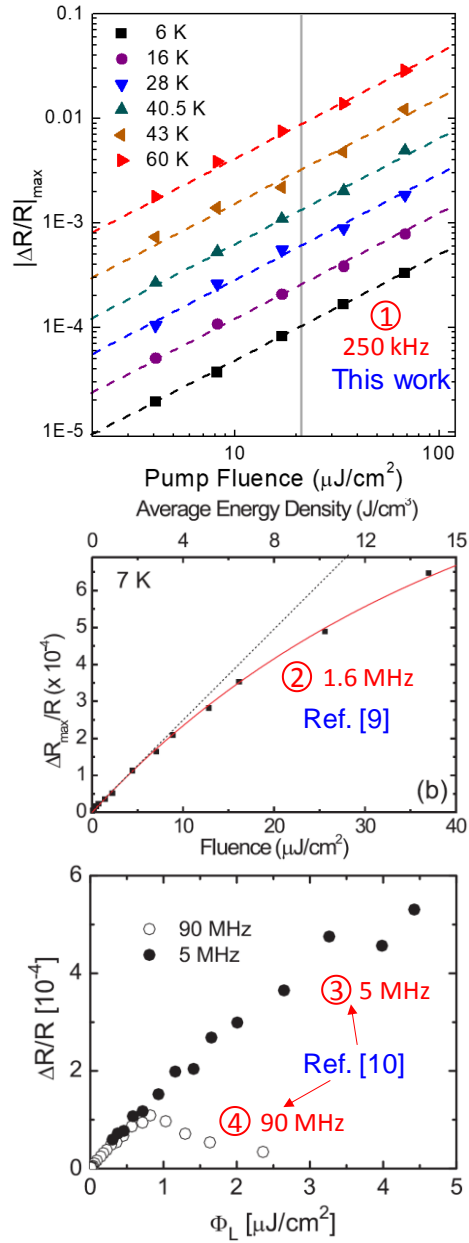

FIG. S5. Fluence threshold of $\Delta R/R_{max}$ for different laser repetition rates. The repetition rate values are in red. (a) Taken from Fig. S5(b). (b) Adapted from ref. [9]. (c) Adapted

from ref. [10]. The threshold values are read when the linear dependence yields to the saturation.

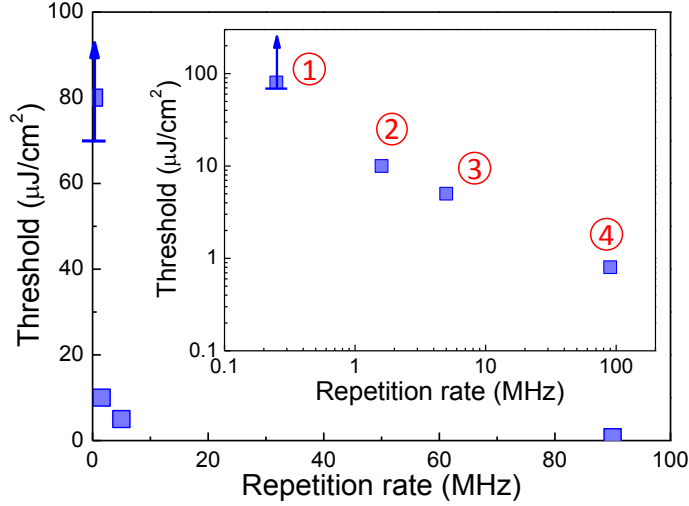

FIG. S6. Fluence threshold depending on the repetition rate. The numbers in the inset correspond to those values in Fig. S5. The threshold reflects possible thermal effect in the material, which is less in an experiment with a lower repetition rate.

### 3.3 Thermal effect in various experiments

We compare the fluence, heat capacity coefficient, and thermal effect among different iron-based superconductors to further ensure that we are in the weak detection regime. In Table S1 we list currently available parameters for iron-based superconductors. The heat capacity coefficient $\kappa$ for $(Li_{0.84}Fe_{0.16})OHFe_{0.98}Se$ is extrapolated from the results in refs [11,12]. Clearly, the fluence in our experiment is on the same order as those in other experiments. Although the fluence is roughly twice of that of bulk FeSe, its heat capacity coefficient is much higher, resulting in a much smaller thermal effect.

The excited-state QPs' temperature can be obtained by (see Section 5)

$$T_e = \langle\sqrt{T_L^2 + \frac{2(1-R)F}{\kappa_v l_s}e^{-z/l_s}}\rangle, \tag{13}$$

where $l_s$ is the optical penetration depth, $R$ is the reflectivity, $\kappa_v = \kappa\rho_n$ is the electronic specific heat capacity coefficient for a unit volume (where $\kappa$ is the electronic specific heat capacity coefficient, and $\rho_n$ is the mole density), $T_L$ is the lattice temperature, and $F$ is the pump fluence. The thermal effect is closely related to the value of $(1-R)F/\kappa$, as listed in Table S1. It can be seen that the thermal effect in our $(Li_{0.84}Fe_{0.16})OHFe_{0.98}Se$ experiment is very small, compared with other reported experiments. Furthermore, the thermal effect value for our $(Li_{0.84}Fe_{0.16})OHFe_{0.98}Se$ is evidently very small. Thus, we conclude the fulfillment of the criterion (A) (*i.e.*, no thermal effect) for our experiment.

Significantly, our laser operates at a repetition rate of 250 kHz, which causes considerably less accumulated thermal heating. In contrast, repetition rates of 1.6 MHz and 5.2 MHz are used in refs [9] and [13], respectively.

Table S1. Pump fluence and thermal effects in different superconductors

| Material | 1UC FeSe | bulk FeSe | $B_{0.6}K_{0.4}Fe_2As_2$ | $(Li_{0.84}Fe_{0.16})OHFe_{0.98}Se$ |
|---|---|---|---|---|
| $F$ ($\mu J/cm^2$) | 268[14] | 9.92[13] | 0.7-36.6[9] | 21 |
| $\kappa$ (mJ·mol$^{-1}$·K$^{-2}$) | 5.73[15] | 5.73[15] | 23[16] | 100[11,12] |
| $(1-R)F/\kappa$ (10 mol K$^2$/m$^2$) | 0.96 | 1.3 | 0.017-0.92 | 0.14 |

## 4. WEAK DETECTION CONDITION: (B) NO PROMINENT DESTRUCTION OF THE SC COMPONENT

### 4.1 No observation of sharp spike in the QP dynamics

For the weak detection criterion (B), an evidence for the occurrence of SC depletion is the observation of a sharp spike nearby $t = 0$. In the below Fig. S7 [8], for a cuprate $Bi_2Sr_2CaCu_2O_{8+\delta}$ at a fluence $\Phi > 70$ $\mu J/cm^2$, a fast peak appears, which "represents the signature of the SC gap collapse" [8]. Throughout our experiment, no such fast spike has been observed. In Fig. S7, the fast spike has a width of 200 fs. As a comparison, our Fig. 1(a) and Figs. S1(e-h) show no such a narrow spike and no such shape a scanning trace. Thus there is no evidence of SC saturation. This is also consistent with the fact that we use a fluence much less than 70 $\mu J/cm^2$, at which the spike starts to appear [8].

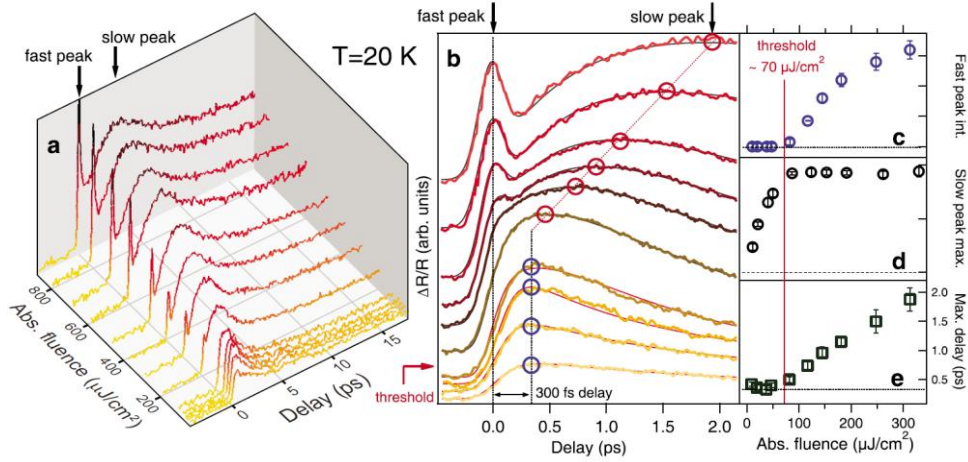

FIG. S7. Evidencing depletion of the SC component (adapted from [8]). Fluence dependence of QP dynamics of $Bi_2Sr_2CaCu_2O_{8+\delta}$. (a) Transient reflectivity traces with different pump fluences. (b) Zoom-in view of panel (a) in the -0.5 – 2.2 ps range. (c-e) fluence dependence of fast peak intensity, slow peak intensity, and delay time of the maximum of the slow peak.

**4.2 Estimation of the fraction of SC electrons that is photo-excited.**

To further verify the fulfillment of the weak detection condition criterion (B), we estimate the fraction of electrons in the SC ground state that are photo-excited and become QPs under laser pulse incidence. Absorbed photons promote the SC ground-state electrons to the excited states, thus breaking the Cooper pairs. However, such electrons only constitute a small portion of the overall SC electrons. We estimate the fraction of such electrons within the sample as follows. Given the pump fluence $F = 21$ μJ/cm$^2$ and the reflectivity $R = 0.31$ (see the discussions in section 5.2), the effective pump fluence is $(1 - R)F$. After passing through one penetration depth for absorption $l_s$ = 25 nm (see the discussions in section 5.1), the pump fluence reduces to $1/e$ times of $(1 - R)F$. The absorbed portion of the pump fluence is thus $(1 - 1/e)(1 - R)F$. Denoting the effective cross-section area of the laser spot in the sample by $S$, the absorbed light energy is $(1 - 1/e)(1 - R)FS$, which occurs within a volume of $l_sS$. Each photon generates a pair of QPs; thus, we determine the density of photo-excited QPs within a unit volume to be

$$\rho_{QP} = 2(1 - 1/e)(1 - R)FS/(\hbar\omega l_s S) = 3 \times 10^{25} m^{-3}. \tag{14}$$

Moreover, the fraction of electrons promoted to the excited state is given as

$$F_{QP} = (\rho_{QP} \times V_{UC})/unit\ cell = 3.9 \times 10^{-3}/unit\ cell, \tag{15}$$

where $V_{UC}$ is the volume of the unit cell and we have taken the lattice parameters $a$, $b$, and $c$ to be 3.7827, 3.7827 and 9.3184 Å [17], respectively. This result clearly shows that only 0.39% of the SC electrons per unit cell are excited by our pump pulses. Hence the SC ground state remains unchanged during the ultrafast dynamics investigation. We conclude that our experiment fulfills very well the weak detection condition criterion (B).

## 5. TEMPERATURE OF PHOTO-EXCITED ELECTRONS $T_e$

Immediately after the pump pulse excitation, the temperature of the photo-excited electrons $T_e$ can be increased to as high as $10^3$ K in a solid, with the lattice temperature $T_L$ still remaining at the ambient temperature. Along the light propagation direction $z$, the energy density $F(z)$ experiences a simple exponential decay due to absorption:

$$F(z) = F_0 e^{-z/l_s}, \qquad (16)$$

where $z$ is the depth away from the sample surface, $l_s$ is the penetration depth, and $F_0 = (1-R)F$ is the energy density right at the surface. Thus, the energy transfer per volume from photons to electrons at position $z$ is $Q(z) = -\frac{dF(z)}{dz} = (1/l_s)F_0 e^{-z/l_s}$. The initial electron temperature is equal to the lattice temperature $T_{ground\ state} = T_L$. Given the linear dependence of electron heat capacity $C_e$ on electron temperature $T_{electron}$, $C_e = \kappa_v T_{electron}$, we have

$$Q(z) = \int_{T_L}^{T_e} C_e dT_{electron} = \int_{T_L}^{T_e} \kappa_v T_{electron} dT_{electron} = (1/2)\kappa_v (T_e^2 - T_L^2), \qquad (17)$$

where $\kappa_v = \kappa \times \rho_n$ is the specific heat capacity coefficient per unit volume, with $\kappa$ being the heat capacity coefficient and $\rho_n$ being the mole density. Thus, we obtain

$$T_e(z) = \sqrt{T_L^2 + \frac{2(1-R)F}{\kappa_v l_s} e^{-z/l_s}}. \qquad (18)$$

As $T_e$ is a function of depth $z$, we consider the average $T_e$ within one penetration depth; thus, we have the expression

$$T_e = \langle \sqrt{T_L^2 + \frac{2(1-R)F}{\kappa_v l_s} e^{-z/l_s}} \rangle. \qquad (19)$$

### 5.1 Penetration depth $l_s$

The penetration depth of some iron-based superconductors has been reported. It is 26 nm for $Ba_{1-x}K_xFe_2As_2$ [9] and 24-26 nm for bulk FeSe [13,18]. These two different types of iron-based superconductors have similar penetration depth. Because (Li$_{0.84}$Fe$_{0.16}$)OHFe$_{0.98}$Se is an intercalated superconductor, which has similar lattice and electronic properties with those of bulk FeSe, we estimate its penetration depth to be similar to those of bulk FeSe and Ba$_{1-x}$K$_x$Fe$_2$As$_2$. As such, we take $l_s$ = 25 nm.

### 5.2 Reflectivity *R*

Measuring the reflectivity *R* of our sample is very challenging because the rough sample surface causes the light to diffuse in all directions, preventing the collection of the total reflected beam. We grow thin film (Li$_{0.84}$Fe$_{0.16}$)OHFe$_{0.98}$Se samples and measure its reflection at low temperatures by using 800 nm femtosecond laser pulses. The transmission of the cryostat window is taken into account. We obtain that *R* = 0.31.

### 5.3 Electron temperature $T_e$

To obtain the electron temperature $T_e$ by Eq. (19), we use a high laser fluence *F*, 2.2 mJ/cm$^2$, and calculate $T_e$ at a lattice temperature of $T_L$ = 7 K. We have estimated that $\kappa$ = 100 mJ·mol$^{-1}$·K$^{-2}$ (Table S1) and $l_s$ = 25 nm. Because in Eq. (19) $T_L^2 \ll \frac{2(1-R)F}{\kappa_v l_s} e^{-z/l_s}$, we have

$$T_e = \left\langle \sqrt{T_L^2 + \frac{2(1-R)F}{\kappa_v l_s} e^{-z/l_s}} \right\rangle \approx (548.5 + \frac{T_L^2}{1074}) \; K \; . \tag{20}$$

With $T_L$ = 7 K, we obtain $T_e$ = 548.5 K. Equation (20) can be plotted in Fig. S8. It can be seen that $T_L$ does not affect $T_e$ very much. At low temperature $T_e$ is nearly a constant.

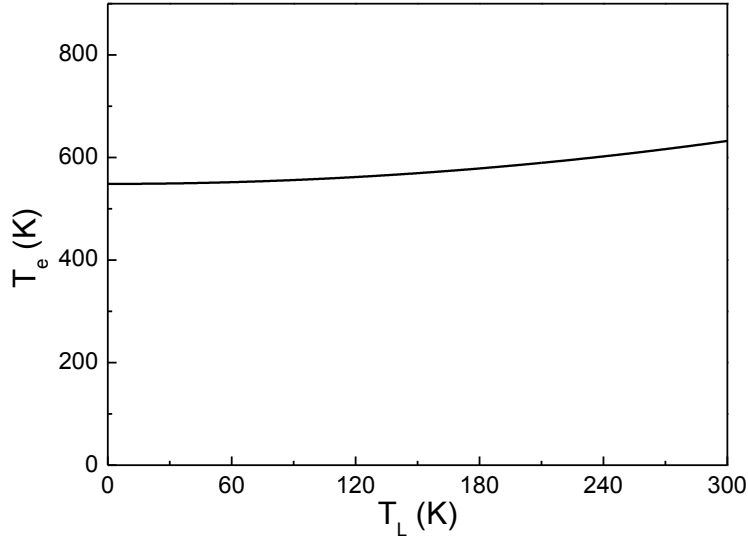

FIG. S8. The $T_L$ dependence of $T_e$ in (Li$_{0.84}$Fe$_{0.16}$)OHFe$_{0.98}$Se. Curve plotted according to Eq. (20). It can be seen that $T_e$ is nearly a constant at a low temperature.

## 6. DECONVOLUTION OF THE QP DYNAMICS

For time-resolved ultrafast spectroscopy experiments, the detected dynamics signal is indeed the convolution between the QP dynamics and the excited-state accumulation response to the laser pulses. This convolution behavior occurs very often. If not correctly de-convoluted, it can lead to fake QP lifetimes in the data analysis.

The detected signal can be expressed as

$$S(t) = D(t) \otimes M(t) = \int_{-\infty}^{+\infty} D(\tau) M(t-\tau) d\tau, \qquad (21)$$

where $S(t)$ is the detected dynamics signal, $D(t) = H(t) A_{\text{fast}} \exp(-t/\tau_{\text{fast}})$ is the fast component of the QP dynamics, $H(t)$ is the unit step function, $A_{\text{fast}}$ is the amplitude of the fast component, $\tau_{\text{fast}}$ is the lifetime of the fast component, and $M(t) = M_0 \exp(-t^2/\tau_M^2)$ is the excited state QP accumulation response to the laser pulse excitation. Although a generalization to include all relaxation components can be implemented, it is mainly the fast relaxation component that is considered here in the deconvolution, because the slow component only slightly convolutes with the response function.

We discuss the response function $M(t)$. Upon light excitation, Cooper pairs are broken to generate QPs, which experience an accumulation process. At time zero, the QPs and HFPs are $n(t = 0)$ and $N(t = 0)$. When the pair-breaking rate $\eta N$ is larger than the QP recombination rate $\beta n^2$, the QP density gradually increases even though the laser pulse has already left (see Eq. (2)) [19]. Thus, the temporal width of the response is larger than the laser pulse duration. We note that not all the accumulation process contributes to the convolution with the QP dynamics. What affects the convolution is the close-to Gaussian response function, which is contained within the accumulation process and cannot be further "reduced". We use a half-Gaussian function to fit the rising accumulation process. Because these accumulated QPs have an identical effect as an ultrafast light pulse, we then extend the response function to a full Gaussian function by adding it rear half (Fig. S9, dashed red curves). Thus we obtain a Gaussian like response function $M(t)$ (Fig. S9). It is this Gaussian response (not the pulse width) that convolutes with the QP dynamics. We note that the effect of the pulse duration is already contained in the response function. As examples, we de-convolute the dynamics for the three FeSe-based superconductors below.

Figure S5 shows the deconvolution for the QP dynamics of single-layer FeSe/SrTiO$_3$ (5 K, replot from ref. [14]), (Li$_{0.84}$Fe$_{0.16}$)OHFe$_{0.98}$Se (7 K, this work), and bulk FeSe (4.4 K, replot from ref. [20]), respectively. Ultrafast laser pulses with temporal durations of 96 fs, 70 fs, and 100 fs are used. In Fig. S9, the QP accumulation Gaussian response function has a temporal width of 490 fs, 125 fs, and 1.30 ps, respectively. In Figs. S9(a-c), the blue curves are the QPs dynamics $D(t)$, the red curves are response functions $M(t)$, and the black curves are the convolution results $S(t)$. The $S(t)$ fits well the experimental data. With such we obtain the fast component lifetimes and ultimately the EPC constants for the three materials (see Fig. S9). We obtain the lifetimes of the fast components as 0.23 ps for single-layer FeSe/SrTiO$_3$, 0.34 ps for (Li$_{0.84}$Fe$_{0.16}$)OHFe$_{0.98}$Se, and 2.04 ps for bulk FeSe. Furthermore, we estimate the EPC strength $\lambda_{A_{1g}}$ to be 0.48 [14], 0.22, and 0.14, respectively.

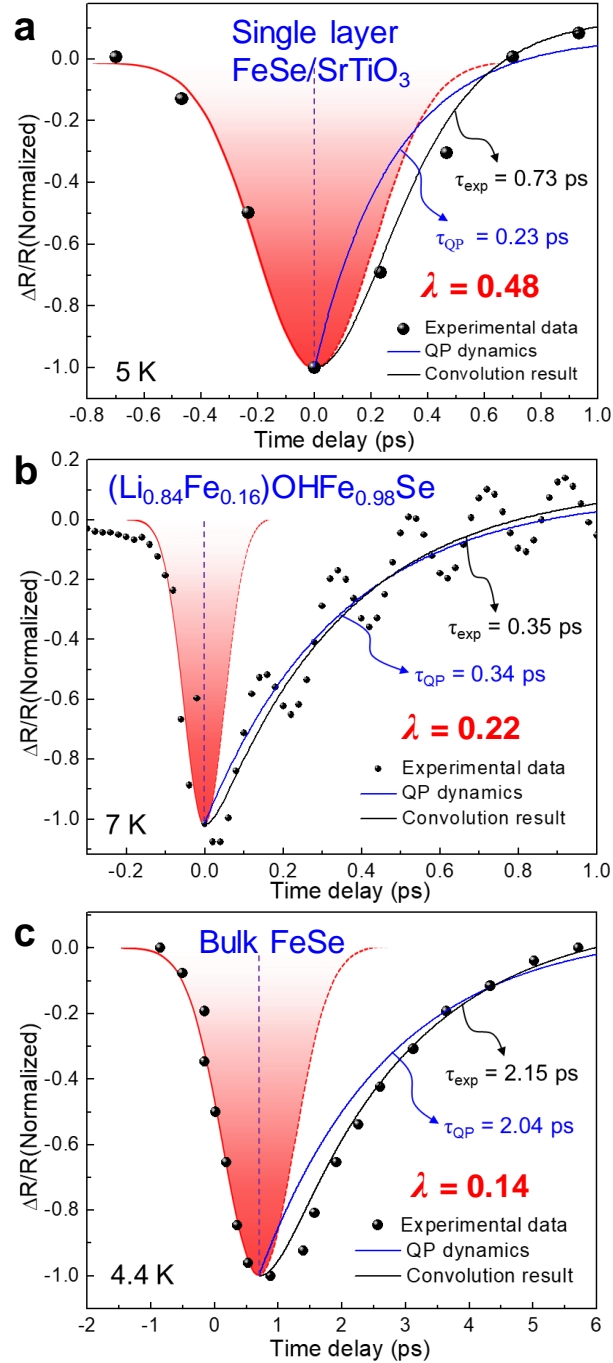

FIG. S9. Deconvolution of the QP dynamics in FeSe-based superconductors. (a) single-layer FeSe/SrTiO$_3$, (b) (Li$_{0.84}$Fe$_{0.16}$)OHFe$_{0.98}$Se, (c) bulk FeSe. Black spheres: experimental data; Black curves: convolution results; Blue curves: QP dynamics; Red curves: response functions $M(t)$, where the dashed curves are the extension to construct a full Gaussian function (see text) and the vertical dashed lines mark the time-zero.

## 7. COHERENT PHONONS

We generate and detect coherent phonons in $(Li_{0.84}Fe_{0.16})OHFe_{0.98}Se$. Because of the multi-crystalline nature of the sample with huge surface light diffusion, the laser fluences used to probe the QP dynamics are too weak to detect the coherent phonons with enough signal-to-noise ratio. Thus, we increase the pump and probe fluences to 2.2 and 0.28 mJ/cm$^2$, respectively. Figure 3a in the main text shows a typical ultrafast dynamics response at 7 K. The periodic oscillations are superimposed on the QP relaxation dynamics, in a manner similar to that observed in other quantum materials [14,21,22]. The frequency of the extracted coherent phonon is 5.11 THz (*i.e.*, 21.2 meV or 171 cm$^{-1}$, see inset in Fig. 3(a)), which can be fitted by a damped cosine function with an initial phase of 0. Thus, the coherent phonon is generated by the displacive excitation of coherent phonons (DECP) mechanism [23]. We assign the phonon to be the $A_{1g}$ mode of Se atoms (as usual, only the $A_1$ mode can be excited in DECP mechanism). The amplitude decreases gradually with increasing temperature (Fig. 3(c)), exhibiting a common anharmonic phonon decay [24]. In contrast, the frequency remains nearly constant with increasing temperature (Fig. 3(d)), demonstrating that no apparent structural phase transition occurs. The insets of Figs. 3(c) and (d) show the magnified view for the 100-180 K temperature range; the temperature dependence in this range is the same as that in the other temperature ranges. No structural transition occurs in this temperature range, which verifies that the carrier density enhancement observed at approximately 120 K in ref. [17] is of electronic rather than lattice origin. Figure 3e shows the phonon lifetime directly measured in the ultrafast spectroscopy experiment, with its temperature dependence. The inset shows the decay rate $\Gamma_\Omega$ of the coherent A$_{1g}$ phonon, which is fitted by $\Gamma_\Omega(T) = \Gamma_0 + \Gamma\left(1 + 2n_{\Omega/2}(T)\right)$ [24,25] (red curve), where $\Omega$ is the phonon frequency, and $n_{\Omega/2}(T) = [\exp(\hbar\Omega/2k_B T) - 1]^{-1}$ is the phonon occupation density. We attribute the phonon decay to the anharmonic phonon interactions.

## 8. OBTAINING THE EPC STRENGTH $\lambda$

In the main text, we stated that $\lambda$ can be obtained by the QP relaxation model [26] under high laser fluence excitation. This way of obtaining $\lambda$ can be applied to various materials, ranging from metals to superconductors. In this model, the QP relaxation rate is determined by

$$\gamma_T = (3\hbar\lambda\langle\Omega^2\rangle/\pi k_B T_e)\left(1 - \frac{\hbar^2\langle\Omega^4\rangle}{(12\langle\Omega^2\rangle k_B^2 T_e T_L)} + \cdots\right), \tag{22}$$

where $\Omega$ is the phonon frequency, $k_B$ is the Boltzmann constant, and $T_e$ is the electron temperature after photo-excitation.

### 8.1 Thermal relaxation rate $\gamma_T$

The thermal relaxation rate $\gamma_T$ reflects the energy transfer from electron to phonons or any other types of elementary excitations, which occurs faster than that associated with phonon-phonon scattering. Thus $\gamma_T$ is related to the initial decay rate in the ultrafast dynamics measurement. We plot the typical scanning traces at different temperatures ranging from 7 K to 290 K in Fig. S10. These traces are obtained with high laser fluence (see the main text for the reason). By focusing on the electronic decay rather than the superimposed oscillations, the QP lifetime at 7 K (Fig. S10(a)) can be obtained to be 0.34 ± 0.05 ps. Thus, $\gamma_T = 1/\tau_{\text{fast}} = 2.9 \pm 0.5$ ps$^{-1}$. We also show the temperature dependence of the QP lifetime in Fig. S10(b). It can be seen that $\tau_{\text{fast}}$ keeps nearly unchanged in the whole temperature range. This result demonstrates that the Allen model is valid in the whole temperature range for $T_L$ (i.e. the higher order terms in Allen model can be neglected).

The requirement for both high temperatures for electron and phonon is due to the *intermediate* mathematical derivation requirement for the Taylor expansion for $N(\Omega,T_L)$ and $N(\Omega,T_e)$, respectively. However, this can be over-tight, because the final result of the Allan model needs only $[N(\Omega,T_L)-N(\Omega,T_e)] \propto [T_L-T_e]$ and the higher terms can be neglected when $T_e T_L$ is relatively high—no requirement on $T_e$ and $T_L$ individually. As discussed in our text (page 6), this is equivalent to "$(E_{phonon})^2 \leq E_{excited\text{-}state\ electron} \cdot E_{ground\text{-}state\ electron}$", which can be fulfilled when $T_e$ is much higher than the optical phonon energy. The experimental results in Fig. S10(b) is an excellent verification and illustration of this reasonable approximation.

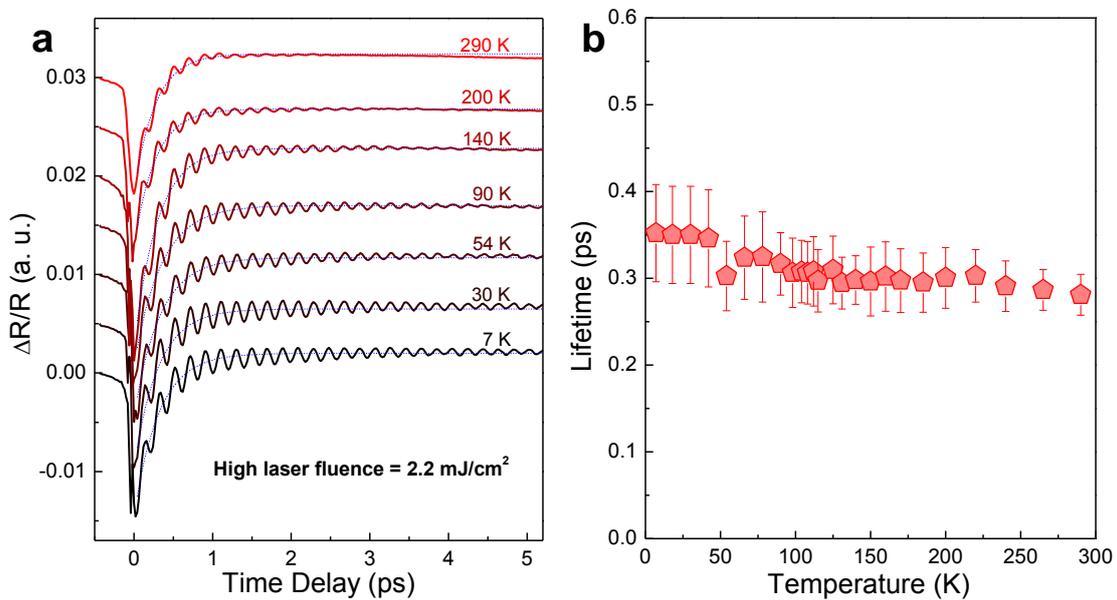

FIG. S10. Temperature dependence of the QP dynamics at 2.2 mJ/cm$^2$. (a) The scanning trace of the QP dynamics at several typical temperatures. The steep increase in the differential reflection $\Delta R/R$ immediately after laser pulse excitation can be used to determine the thermal relaxation rate $\gamma_T$. The oscillations superimposed on the exponential QP dynamics correspond to the coherent phonon. The blue short-dashed curves are the single-exponential fitting curves for the electronic ultrafast dynamics (fast component). (b) The temperature dependence of QP lifetime.

A more precise way to obtain the $\gamma_T$ is through the two-temperature model (TTM) described by the thermodynamics differential equations [26,27]. The exponential fitting method we use here yields equivalently accurate results, as shown in Fig. S11. In deriving the TTM *differential* result, we analyze the *difference* results for delay times. Upon approaching time zero, this yields an ideal value of $\gamma_T$, which compares excellently with the exponential fit result. This is mainly because the exponential function can fit our time-resolved experimental data very well.

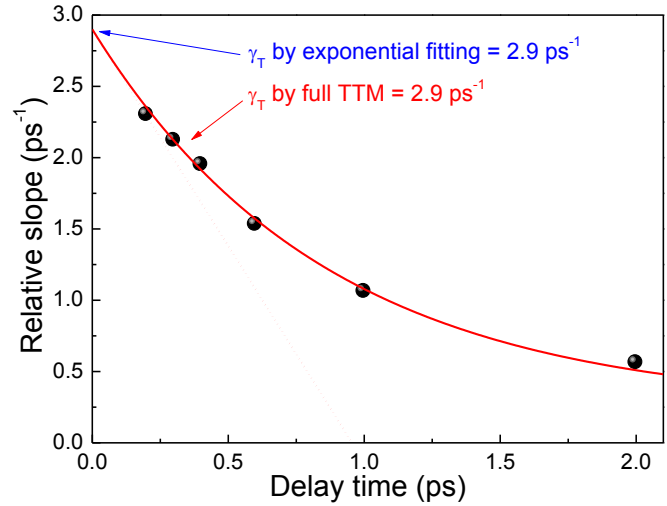

FIG. S11. Comparison of $\gamma_T$ obtained by full TTM and by exponential fitting. Black sphere: slope obtained at the given delay time. Red solid curve: guide to the eyes for the trend of variation. Red dashed line: asymptotic line.

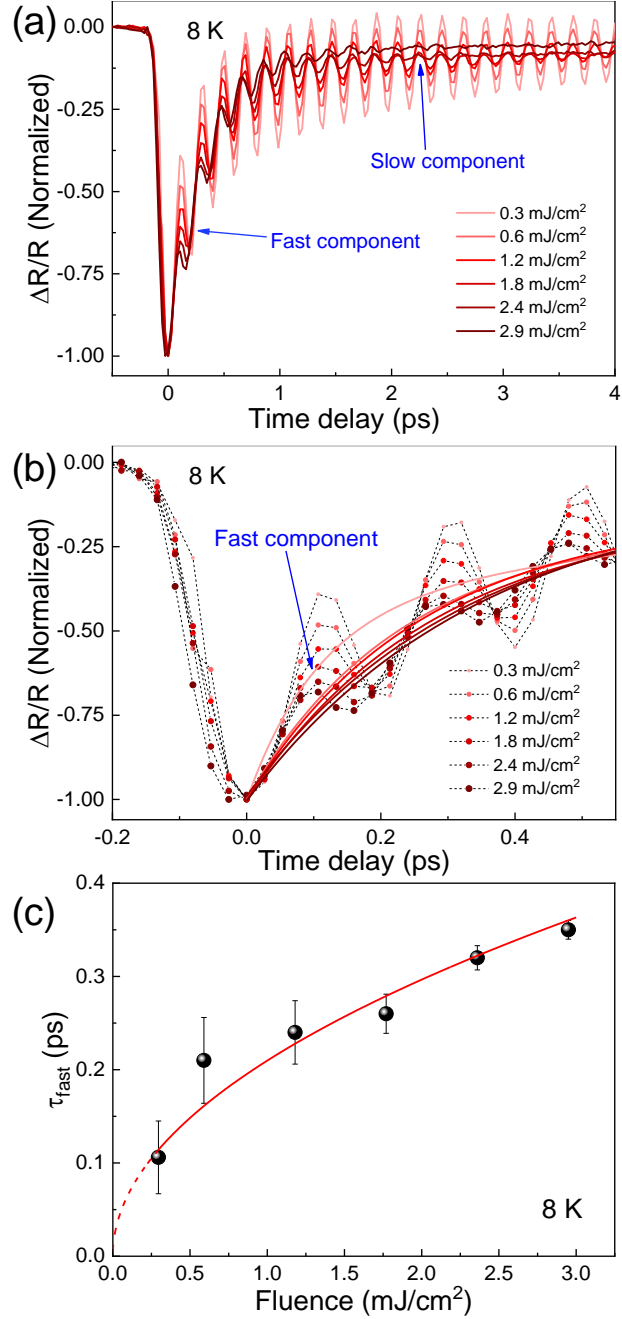

FIG. S12. Fluence dependence of $\tau_{fast}$ in the high fluence regime. (a) Normalized fluence-dependent QP dynamics at 8 K. (b) Zoom-in view of the fast component shown in (a). The solid curves are fittings to the dynamics other than the coherent oscillations. (c) Fluence dependence of $\tau_{fast}$ derived from (b). The dashed curve is an extension of the solid fitting curve (see text of SM).

We also did the fluence dependence experiment and the result is shown in Fig. S12.

In Fig. S12(a) we show the normalized dynamics results, each of which contains two components (a fast one and a slow one). To see the fast component clearly, we re-plot a zoom-in view of the short-time region of Fig. S12(a) in Fig. S12(b). It can be seen that the fast component decays slower at higher fluences. Quantitatively, in Fig. S12(c) we summarize the lifetimes $\tau_{fast}$ as a function of fluence, which can be well fitted by a relation $\tau_{fast} = \frac{\pi k_B}{3\hbar \langle \omega^2 \rangle \lambda} \sqrt{T_l^2 + \frac{2(1-R)F}{\kappa_v l_s} e^{-z/l_s}}$. This relation is derived from $\lambda = \pi k_B T_e / \left(3\hbar \langle \omega^2 \rangle \tau_{fast}\right)$ (see Ref. [26]). As shown in Fig. S12(c), the estimated $\tau_{fast}$ does vary as a function of fluence, demonstrating that $\lambda$ is a constant in the high-fluence regime.

### 8.2 The EPC strength $\lambda$

Given all the parameters above, we can now obtain the EPC strength $\lambda$ under the condition of low lattice temperature and high laser fluence. We emphasize in the main text that the higher terms in Eq. (16) can be neglected when $(E_{phonon})^2 \leq E_{excited\text{-}state\ electron} \cdot E_{ground\text{-}state\ electron}$ (*i.e.*, the energy of the excited-state electron is much higher than that of the phonon). Note that $\lambda$ is a constant, which is a fundamental property that does not rely on laser fluence and persists up to a temperature much higher than the SC $T_c$. As a deduction, $\gamma_T$ is expected to be smaller at higher $T_e$, which has been verified in Ba(Fe$_{0.92}$Co$_{0.08}$)$_2$As$_2$ [28]. As stated in the main text, we mainly consider the A$_{1g}$ mode phonon, whose frequency is 5.11 THz (*i.e.*, 21.2 meV or 171 cm$^{-1}$). By Eq. (22), we obtain that $\lambda_{A_{1g}} = 0.22 \pm 0.04$, which is plotted in Fig. 4 of the main text.

### 9. THE $T_c$ AND NOMINAL $\lambda_{A_{1g}}$ VALUES SHOWN IN FIG. 4

In our experiment, we measure the quantity $\langle \lambda \Omega^2 \rangle$. In Allen-Dynes' treatment,

the logarithmic average phonon frequency $\Omega_{\log}$ is used ("with a prefactor altered from $\Theta_D/1.45$ to $\omega_{\log}/1.2$") to make the McMillan equation "highly accurate" [abstract, ref. 29]. As a reasonable treatment, we assign $\langle\lambda\Omega^2\rangle \approx \langle\lambda\Omega_{\log}^2\rangle = \lambda\Omega_{\log}^2$. Consequently, a natural and appropriate definition of the nominal $\lambda_{A1g}$ is $\lambda_{A1g} = \lambda(\Omega_{\log}^2/\Omega_{A1g}^2)$, as given in the main text (page 8).

As shown in Fig. 4 of the main text, all categories of FeSe- and FeAs-based superconductors are considered (mostly nearly optimized materials). Here, we explicitly list their $T_c$ and $\lambda$ values in Table S2, along with the corresponding references.

**Table S2.** The $T_c$ and $\lambda_{A_{1g}}$ values of iron-based superconductors shown in Fig. 4

| Superconductors | $T_c$ (K) | $\lambda_{A_{1g}}$ |
|---|---|---|
| Bulk FeSe | 8[30] | 0.07[retrieved from 31] |
| Bulk FeSe | 8[30] | 0.16[13] |
| NaFeAs | 9[32] | 0.1[retrieved from 33] |
| $Fe_{1.05}Se_{0.2}Te_{0.8}$ | 10[this work] | 0.11 ±0.03[this work] |
| $Fe_{1.01}Se_{0.2}Te_{0.8}$ | 13.5[this work] | 0.13 ±0.03[this work] |
| $FeSe_{0.5}Te_{0.5}$ | 14.5 | 0.13[retrieved from 34] |
| LiFeAs | 18[35] | 0.12[retrieved from 33] |
| $CaFe_{1.85}Co_{0.15}As_2$ | 20[36] | 0.115[retrieved from 36] |
| $BaFe_{1.85}Co_{0.15}As_2$ | 23[37] | 0.125[37] |
| $BaFe_{1.84}Co_{0.16}As_2$ | 24[28] | 0.12[28] |
| $BaFe_{1.85}Co_{0.15}As_2$ | 25[38] | 0.16[38] |
| $LaFe_{1.85}AsO_{0.875}F_{0.125}$ | 26[2] | 0.16[39] |
| $K_xFe_2Se_2$ | 30[40] | 0.175[retrieved from 41] |
| $Ba_{0.6}K_{0.4}Fe_2As_2$ | 38[42] | 0.207[retrieved from 42] |
| $(Li_{0.84}Fe_{0.16})OHFe_{0.98}Se$ | 39.7 ±0.5[this work] | 0.22 ±0.04[this work] |
| 1UC $FeSe/SrTiO_3$ | 58 ±7[43] | 0.5[43] |
| 1UC $FeSe/SrTiO_3$ | 68(-5/+2)[14] | 0.48[14] |

A few important notes and explanations for obtaining the values in Table S2 are given below.

### 9.1 Obtaining the experimental values of $\lambda_{A_{1g}}$ shown in Fig. 4

The $T_c$ values of the single-layer FeSe/SrTiO$_3$ samples [14,43] are lowered by their protecting FeTe layers or their removal. Thus indeed the $T_c$ values could be slightly higher. Thus the error bars shown in Fig. 4 are only nominal error bars for the two results. The $\lambda$ value of BaFe$_{1.85}$Co$_{0.15}$As$_2$ in ref. [37] are not directly given by the authors of ref. [37]; we extract the ultrafast dynamics result from ref. [37], perform deconvolution, obtain the QP decay lifetime, and calculate the $\lambda_{A_{1g}}$ to be 0.125 (using the A$_{1g}$ phonon frequency value given in ref. [37]). Note that both [37] and our treatment here of the reported data share an identical method of obtaining $\lambda_{A_{1g}}$ as in ultrafast optical spectroscopy; hence we classify [37] into the ultrafast experiment category in Fig. 4. The lateral error bar for the bulk FeSe is given not by the authors of ref. [13]; we extract the ultrafast dynamics result from ref. [13], perform deconvolution, obtain the QP decay lifetime to be 2.04 ps, and derive the $\lambda_{A_{1g}}$ to be 0.14, which we set to be the lower limit of the data fluctuation. It is worthy to note that the SC properties are very much sensitive to the sample-to-sample fluctuation for bulk FeSe. The sphere in golden color is a result not obtained by using the A$_{1g}$ phonon frequency. All the other data points are obtained by using the A$_{1g}$ phonon frequency.

Furthermore, we performed to additional experiments on two iron-based superconductors, respectively. We grow two nearly optimized samples: Fe$_{1.05}$Te$_{0.8}$Se$_{0.2}$ ($T_c$ = 10 K) and Fe$_{1.01}$Te$_{0.8}$Se$_{0.2}$ ($T_c$ = 13.5 K). The pump fluence is 16 $\mu$J/cm$^2$, which is nearly identical to that in ref. [20], where the photon energy is 1.55 eV for pump and probe beam. We measure the dynamics at 10 K in the time delay range of -0.5 – 110 ps, as shown in Fig. S13. The fast component with a few hundreds of femtosecond lifetime is clearly detected on both two samples. For $Fe_{1.05}Te_{0.8}Se_{0.2}$, the fast component lifetime is 0.31 ±0.05 ps, and for $Fe_{1.01}Te_{0.8}Se_{0.2}$ the fast component lifetime is 0.27 ±0.05 ps. By using Allen model and comparing with ref. [20], we identify the $\lambda_{A1g}$ value of $Fe_{1.05}Te_{0.8}Se_{0.2}$ and $Fe_{1.01}Te_{0.8}Se_{0.2}$ to be 0.11 ±0.03 and 0.13 ±0.03, respectively.

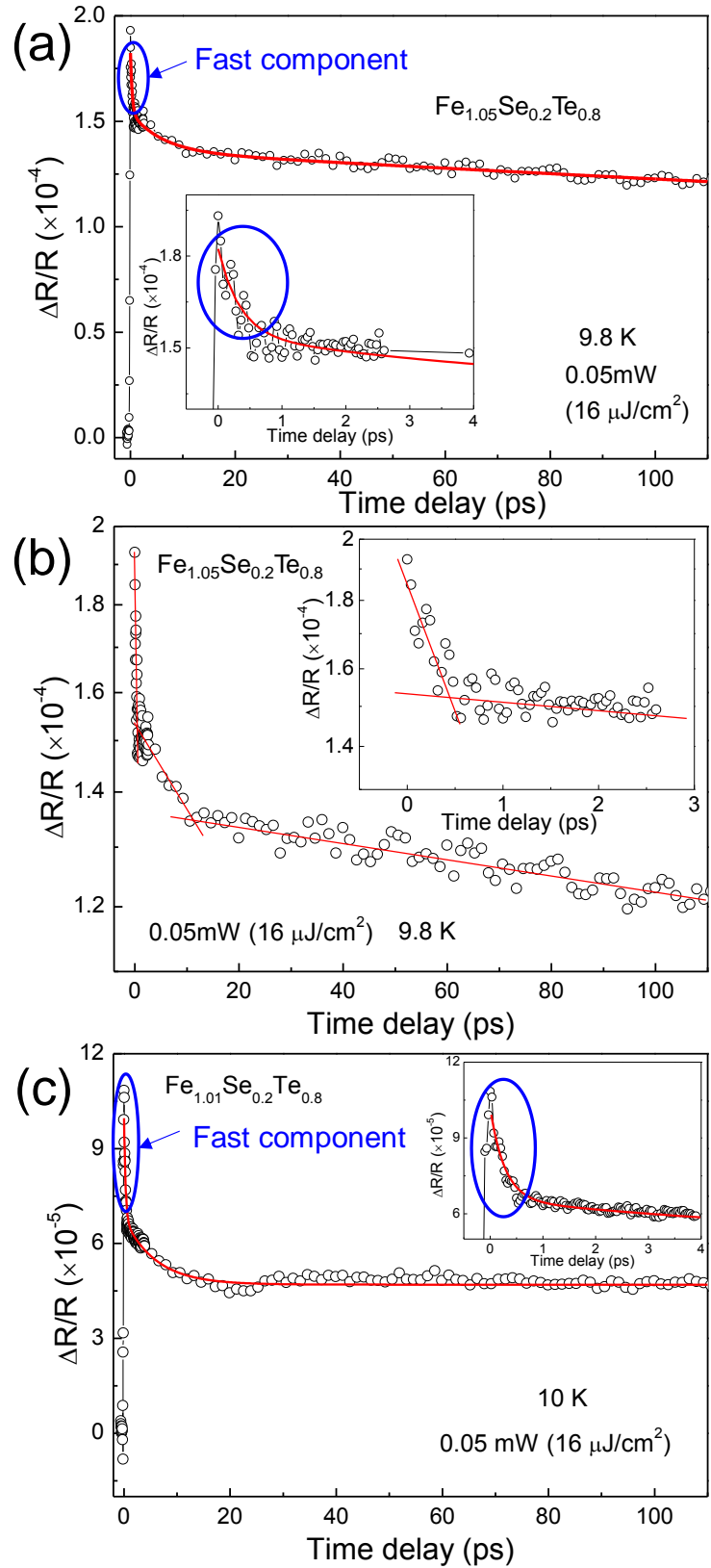

FIG. S13. Ultrafast dynamics of $Fe_{1.05}Te_{0.8}Se_{0.2}$ and $Fe_{1.01}Te_{0.8}Se_{0.2}$. (a) The scanning trace of $Fe_{1.05}Te_{0.8}Se_{0.2}$ at 9.8 K. (b) The y-axis logarithmic scale plot of the same data

shown in (a). The red solid lines are guides to the eyes for the multi-components of the dynamics. (c) The scanning trace of $Fe_{1.01}Te_{0.8}Se_{0.2}$ at 10 K. The red solid curves in (a) and (c) are fitting curves with multi-exponential functions. The inset figures are zoom-in views. The fast components are marked by blue circles.

### 9.2 Retrieving the theoretical values of $\lambda_{A_{1g}}$ shown in Fig. 4

In Fig. 4 of the main text, the $\lambda_{A_{1g}}$ values of LiFeAs [33], NaFeAs [33], $CaFe_{1.85}Co_{0.15}As_2$ [36], $K_xFe_2Se_2$ [41], $Ba_{0.6}K_{0.4}Fe_2As_2$ [42], and $FeSe_{0.5}Te_{0.5}$ [34] superconductors were not reported by their authors, but are retrieved from the references by us. In these references, the theoretical $\lambda$ values are calculated based on the Eliashberg function using the logarithmic average phonon frequency $\Omega_{\log}$ (Table S3). The average phonon frequency $\Omega_{\log}$ and the $A_{1g}$ phonon frequency $\Omega_{A_{1g}}$ were also reported (Table S3). Given that $\lambda \propto 1/\langle\Omega^2\rangle$ [44], to the first order of approximation, we retrieve the values of $\lambda_{A_{1g}}$ by using $\lambda_{A_{1g}} = \lambda(\Omega_{\log}^2/\Omega_{A_{1g}}^2)$.

For LiFeAs and NaFeAs, their $\Omega_{\log}$ values have not been reported. We estimate it by the following way. We illustrate the $\Omega_{A_{1g}}$ vs $\Omega_{\log}$ of Fe-based superconductors in below Fig. S14. It can be seen that the $\Omega_{A_{1g}}$ and $\Omega_{\log}$ values have a positive correlation (the yellow stripe). We assume that LiFeAs and NaFeAs also obey this correlation. From the plot, we can obtain the $\Omega_{\log}$ of LiFeAs is ($120 \pm 15$ cm$^{-1}$) and that of NaFeAs is $100 \pm 15$ cm$^{-1}$. Thus the corresponding $\lambda_{A_{1g}}$ values are obtained to be 0.12 and 0.1, respectively.

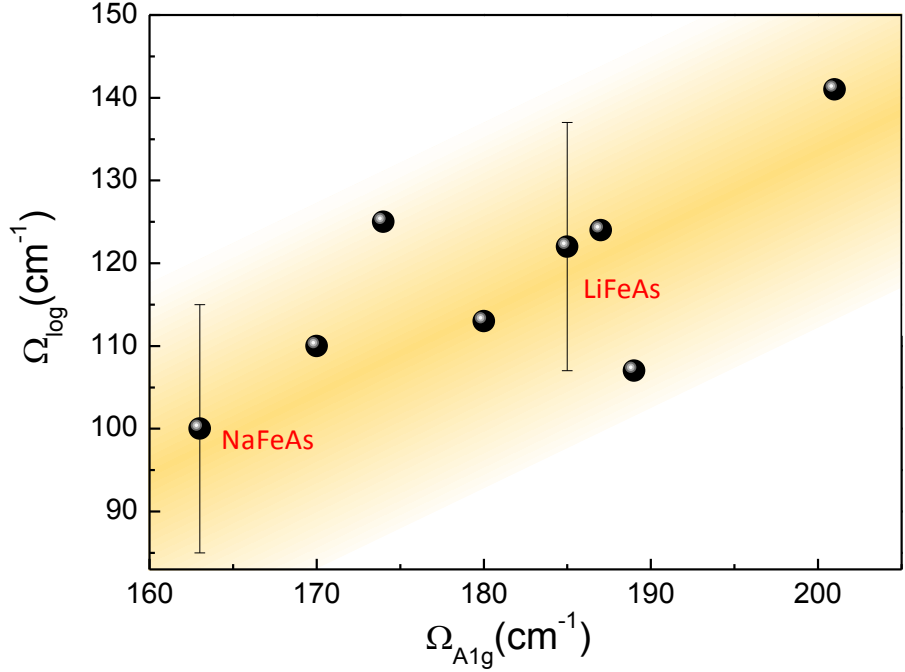

FIG. S14. Estimation of the $\Omega_{\log}$ value of LiFeAs and NaFeAs. Yellow stripe: overall positive correlation between $\Omega_{A1g}$ and $\Omega_{\log}$.

Thus, we obtain the values of $\lambda_{A_{1g}}$ (Table S3). We summarize these retrieval processes in Fig. 4, where the original reported $\lambda$ values are marked by light gray squares or spheres, and white dashed arrows point from the reported $\lambda$ values to the retrieved $\lambda_{A_{1g}}$ values. These retrieved theoretical $\lambda_{A_{1g}}$ values and the reported experimental $\lambda_{A_{1g}}$ values together form a whole set of data shown in Fig. 4, which fall within the range of the purple stripe (centered on the pink curve, indicating a positive correlation). We emphasize that all the cited theoretical works use the same definition of $\lambda$ following Allen, and the experimental data all rely on the ultrafast dynamics described by the Allen model (except [43]).

The pink curve is a modified Allen-Dynes formula, where $\lambda$ is replaced by $6.5\lambda_{A_{1g}}$, and $\Omega_{\log}$ by $1.5\Omega_{A1g}$:

$$T_c = 1.5 \frac{\Omega_{A_{1g}}}{1.2} \exp\left[-\frac{1.04\left(1+6.5\lambda_{A_{1g}}\right)}{6.5\lambda_{A1g}-\mu^*\left(1+0.62\times 6.5\lambda_{A1g}\right)}\right] \quad (23)$$

To the first order of approximation, we specify an average value of $\Omega_{A1g}$ for the various materials. We choose 176 cm$^{-1}$ (i.e., 254 K), where in Fig. 4 the smallest value is 163 cm$^{-1}$ (for NaFeAs) and the largest is 189 cm$^{-1}$ (for CaFe$_{1.85}$Co$_{0.15}$As$_2$).

The orange curve is another modified Allen-Dynes formula, with $\lambda$ replaced by $2.7\lambda$ and $\Omega_{\log}$ replaced by $2.31\Omega_{\log}$:

$$T_c = 2.31 \frac{\Omega_{\log}}{1.2} \exp\left[-\frac{1.04(1+2.7\lambda)}{2.7\lambda-\mu^*(1+0.62\times 2.7\lambda)}\right] \quad (24)$$

Similarly, we specify an average value of $\Omega_{\log}$ to be 120 cm$^{-1}$ (i.e., 173 K).

**Table S3.** Retrieval of $\lambda_{A_{1g}}$ from the theoretical values of $\lambda$

| Material | $\Omega_{\log}$ (cm$^{-1}$) | $\lambda$ | $\Omega_{A1g}$ (cm$^{-1}$) | $(\Omega_{\log}/\Omega_{A1g})^2$ | $\lambda_{A1g}$ | $T_c$(K) |
|---|---|---|---|---|---|---|
| FeSe | 113[31] | 0.17[31] | 180[45] | 0.394 | 0.067 | 8 |
| NaFeAs | 100 | 0.27[33] | 163[46] | 0.376 | 0.1 | 9 |
| FeSe$_{0.5}$Te$_{0.5}$ | 110[34] | 0.3[34] | 170 | 0.419 | 0.13 | 14.5 |
| LaFe$_{1.85}$AsO$_{0.875}$F$_{0.125}$ | 141[47] | 0.35 | 201[48] | 0.46 | 0.16[39] | 26 |
| CaFe$_{1.85}$Co$_{0.15}$As$_2$ | 107[36] | 0.36[36] | 189[49] | 0.321 | 0.115 | 16 |
| LiFeAs | 120 | 0.29[33] | 185[50] | 0.42 | 0.12 | 18 |
| K$_x$Fe$_2$Se$_2$ | 125[41] | 0.34[41] | 174[51] | 0.516 | 0.175 | 30 |
| Ba$_{0.6}$K$_{0.4}$Fe$_2$As$_2$ | 124[42] | 0.47[42] | 187[52] | 0.44 | 0.207 | 38 |

# 10. AFFECTING THE EPC THROUGH MODIFING THE ELECTRONIC STATES, PHONONS, OR BOTH

Our results suggest that all iron-based superconductors have the same SC mechanism, as reflected in Fig. 4. These data include almost all types of bulk iron-based superconductors, including the ones we are measuring now. This surprising result in Fig. 4 suggests a correlation between EPC and superconductivity for iron-based superconductors. The physical picture might be like this (Fig. S15): the EPC plays a crucial role in the SC pairing, while various other types of excitations, interactions or modifications may affect either the electrons or the phonons or both; as a result, the superconducting properties are largely affected.

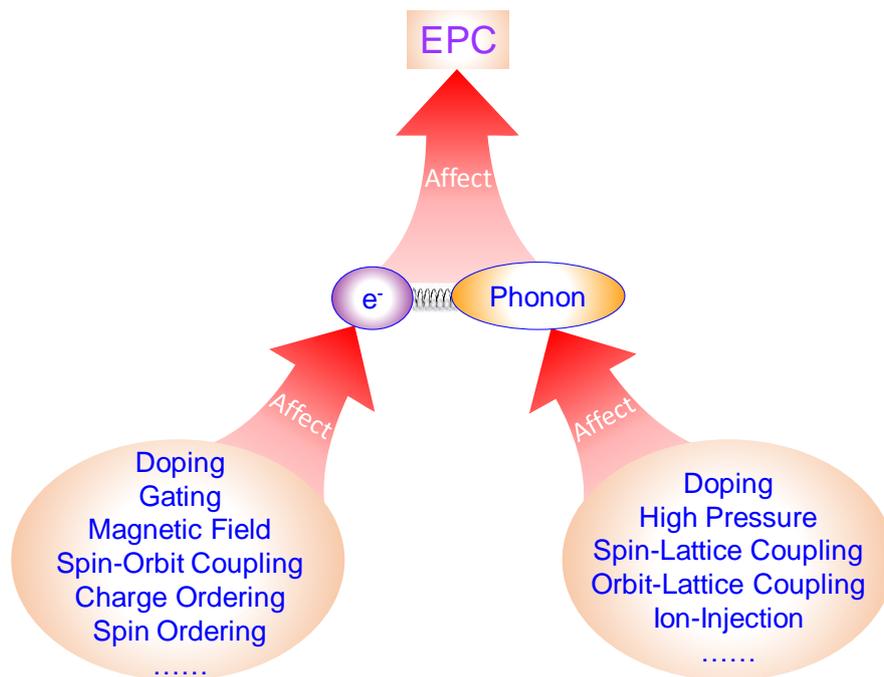

FIG. S15. Schematics of the factors that can affect the EPC through affecting the electrons or phonons or both. The EPC can be affected by modifications in the electronic states, phonons, or both. The electronic states can be affected by doping, gating, magnetic field, spin ordering, charge ordering, spin-orbit coupling, *etc.*; the phonons can be affected by doping, high pressure, spin-lattice coupling, orbit-lattice coupling, ion injection, *etc.*